\documentclass[preprintnumbers,twocolumn,article,amsmath,amssymb,floatfix,9pt,prd,superscriptaddress,nofootinbib]{revtex4-1}

\usepackage[utf8,latin1]{inputenc}
\usepackage{graphicx}
\usepackage{dcolumn}
\usepackage[T1]{fontenc}

\usepackage{mathrsfs}
\usepackage{cases}
\usepackage{bm}
\usepackage{academicons}
\usepackage{mathtools,nccmath}
\usepackage{fancyhdr}
\usepackage{tikz}
\usepackage[usenames,dvipsnames]{xcolor}
\definecolor{fixorange}{RGB}{255,140,0}

\usepackage{tensor}
\usepackage[normalem]{ulem}
\usepackage{lipsum}
\usepackage{soul}
\usepackage{cancel}
\usepackage{stackengine,scalerel}
\usepackage{mathabx}
\usepackage{hyperref}
\usepackage{tabularx}
\usepackage{makecell}
\usepackage{booktabs}

\hypersetup{colorlinks,linkcolor={red},citecolor={blue},urlcolor={blue}}

\AtBeginDocument{\fontsize{8}{11}\selectfont}

\renewcommand{\arraystretch}{1.2}

\definecolor{lime}{HTML}{A6CE39}
\DeclareRobustCommand{\orcidicon}{%
	\begin{tikzpicture}
	\draw[lime, fill=lime] (0,0)
	circle [radius=0.16]
	node[white] {{\fontfamily{qag}\selectfont \tiny ID}};
	\draw[white, fill=white] (-0.0625,0.095)
	circle [radius=0.007];
	\end{tikzpicture}
	\hspace{-2mm}
}

\fancypagestyle{plain}{%
  \fancyhf{}
  \fancyfoot[C]{\iffloatpage{}{\thepage}}
  }
\foreach \x in {A, ..., Z}{%
	\expandafter\xdef\csname orcid\x\endcsname{\noexpand\href{https://orcid.org/\csname orcidauthor\x\endcsname}{\noexpand\orcidicon}}
}

\begin{document}

\title[Rotating Barrow entropy-inspired wormholes versus black holes:\\
null trajectories, shadow morphology, and exotic matter]{Rotating Barrow entropy-inspired wormholes versus black holes: \\null trajectories, shadow morphology, and exotic matter}

\author{A. Errehymy\orcidA{}}
\email{abdelghani.errehymy@gmail.com }
\affiliation{Astrophysics Research Centre, School of Mathematics, Statistics and Computer Science, University of KwaZulu-Natal, Private Bag X54001, Durban 4000, South Africa}
\affiliation{Center for Theoretical Physics, Khazar University, 41 Mehseti Str., Baku, AZ1096, Azerbaijan}
%\affiliation{Jadara Research Center, Jadara University, Irbid 21110, Jordan}

\author{G. Yergaliyeva\orcidB{}}
\email{gyergaliyeva1171@gmail.com}
\affiliation{Department of Natural Sciences and Pedagogical Innovations, L.N. Gumilyov Eurasian National University, Astana, 010008, Kazakhstan}

\author{S. K. Maurya \orcidJ{}}
\email{sunil@unizwa.edu.om}
\affiliation{Department of Mathematical and Physical Sciences, College of Arts and Sciences, University of Nizwa, Nizwa 616, Sultanate of Oman}
%\affiliation{Jadara Research Center, Jadara University, Irbid 21110, Jordan}

\author{B.~Turimov\orcidD{}}
\email[]{bturimov@astrin.uz}
\affiliation{Engineering School, Central Asian University, Milliy bog Str.264, Tashkent, 111221, Uzbekistan}      
\affiliation{University of Tashkent for Applied Sciences, Gavhar Str.1, Tashkent, 100149, Uzbekistan}
%\affiliation{Ulugh Beg Astronomical Institute, Astronomy St. 33, Tashkent 100052, Uzbekistan}

\author{I.~Ibragimov \orcidD{}}
\email{i.ibragimov@kiut.uz}
\affiliation{Kimyo International University in Tashkent, Shota Rustaveli Street 156, Tashkent 100121, Uzbekistan}

\author{J. Rayimbaev\orcidE{}}
\email{javlonrayimbaev6@gmail.com}
\affiliation{School of Physics, Harbin Institute of Technology, Harbin 150001, China}
\affiliation{Institute of Theoretical Physics, National University of Uzbekistan, Tashkent 100174, Uzbekistan}
%\affiliation{Tashkent State Technical University, Tashkent 100095, Uzbekistan}

\date{\today}

\begin{abstract}
 We construct a rotating traversable wormhole supported by a regular, entropy-inspired matter distribution and investigate its geometry, null geodesics, shadow morphology, and exotic matter content. We derive the geometry from the Barrow deformation of the Bekenstein--Hawking area law, characterized by an exponent $0<n\le1$, which replaces point-like divergences with a regular matter profile. The resulting energy density is negative everywhere and scales as $\rho_B\propto-r^{-n-3}$, while the shape function smoothly satisfies the flare-out condition at the throat. We introduce slow rotation via $N(r,\theta)=\exp[\alpha J^2\cos^2\theta/r^2]$ and $\omega(r)=2J/r^3$; both remain finite at the throat, ensuring a horizon-free, traversable geometry. Equatorial null geodesics yield a photon-sphere condition distinct from the standard black-hole relation. In the static limit, the photon sphere sits at the throat, $r_{\rm ph}=r_{\rm th}$; for slow rotation, the prograde branch moves to $r_{\rm ph}^2\simeq 6J$, while the retrograde branch is pushed toward the throat. The critical impact parameters are $\mathscr{B}_{\pm}\approx\pm r_{\rm ph}+2J/r_{\rm ph}$. For $M=1$, $r_{\rm th}=1$, $J/M^2=0.5$, this gives $r_{\rm ph}\approx1.732$, $\mathscr{B}_{+}\approx1.90$, $\mathscr{B}_{-}\approx-1.55$, and a fractional asymmetry $\mathcal{A}\approx0.10$. The weak and null energy conditions are violated, with the strongest violation near the throat, while the Barrow exponent controls the localization of exotic matter. The volume-integral quantifier $I_V$ remains finite when $r_{\rm th}^{n+1}=4M/[(n+2)\pi^{n/2}]$. Comparisons with Kerr and regular black holes indicate that shadow diameter alone may be degenerate, whereas photon-ring structure and rotational asymmetry provide complementary signatures capable of distinguishing this entropy-inspired wormhole from black-hole alternatives.\\
\keywords{Rotating traversable wormholes, Barrow entropy, black hole shadows}
\end{abstract}

\maketitle

\section{Introduction}\label{Sec:I}
%---------------------------------------------

A wormhole can be viewed as a geometric shortcut connecting distant regions of spacetime. The idea dates back to Einstein and Rosen, who showed in 1935 that the Schwarzschild solution can be interpreted as a bridge connecting two asymptotically flat regions~\cite{Einstein:1935}. For several decades, however, this construction remained largely a theoretical curiosity. This perspective changed with Morris and Thorne~\cite{Morris:1988}, who asked a more practical question: what physical conditions keep a wormhole throat open and traversable for a traveler? They found that a traversable wormhole must be horizon-free and that its supporting stress--energy tensor must violate the classical energy conditions. In particular, the radial tension required to sustain the throat must exceed the local energy density, a property ordinary matter does not exhibit under standard physical conditions. This requirement has given rise to the long-standing problem of identifying or constructing suitable exotic matter capable of supporting traversable wormholes.

Since then, much effort has gone into making that requirement less embarrassing. Visser showed that the offending material can be squeezed onto a thin shell at the throat, which at least localizes the pathology~\cite{Visser:1989,Visser:1995}. Hochberg and Visser then sharpened the geometric side of the story, deriving the general flare-out condition that any static throat must satisfy~\cite{Hochberg:1998}. The volume-integral quantifier introduced by Visser, Kar, and Dadhich gave a practical way to measure how much exotic matter a given geometry actually needs, with the mildly surprising conclusion that the total can be made arbitrarily small~\cite{Visser:2003}. Lobo and collaborators examined a wide range of concrete sources, including phantom scalars, nonminimal curvature couplings, and related constructions, to assess which of them could plausibly support wormhole geometries~\cite{Lobo:2009,Garcia:2011}. A parallel line of work shifted the discussion to modified gravity, where extra geometric terms can mimic the required stress--energy without introducing fundamentally unphysical matter~\cite{Bisabr:2012tg,Mehdizadeh:2015,Ibadov:2020btp,Bambi:2013,Asghar:2026lll}. Embedding wormholes in dark matter halos is another workable route, since the surrounding medium can supply part of the negative pressure~\cite{Errehymy:2024lhl,Guvendi:2025gej,Errehymy:2025vvs,Errehymy:2025kzj,Errehymy:2026pbo}. None of these constructions matters much astrophysically unless the wormhole can spin. Real compact objects rotate, and rotation is not a small perturbation to be waved away: it drags inertial frames, distorts the photon region, and imprints a left--right asymmetry on anything a distant observer sees. Teo wrote down the general stationary and axisymmetric wormhole metric and, in doing so, gave the field a working framework for rotating solutions~\cite{Teo:1998}. The follow-up literature is broad. Researchers have studied rotating wormholes in modified gravity~\cite{Kumar:2023wfp,Garattini:2025cix,Errehymy:2026xoh,Uemichi:2026dzb}, noncommutative settings~\cite{Errehymy:2026mok}, and with a wide range of matter sources. Their shadows and photon rings have only recently received the attention they deserve, largely because of the observational developments described below~\cite{Nedkova:2013msa,Ohgami:2015nra,Saleem:2024,Lutfuoglu:2026gey}.

The Event Horizon Telescope changed the tone of this discussion. The images of M87$^*$ and Sgr A$^*$ are not merely pretty pictures; they are the first direct measurements of horizon-scale structure, and they show a ring of emission surrounding a central brightness depression~\cite{EHT:2019,EHT:2022}. The size and shape of that depression, the position and width of the photon ring, and the spin-induced asymmetry of the whole image all carry information about the metric near the compact object~\cite{Johannsen:2013,Broderick:2022,Gralla:2020,Johnson:2020,Narayan:2019,Tao:2023hou}. The Kerr metric remains the default interpretation, but it is no longer the only option. Modified gravity, quantum corrections, and horizonless compact objects each predict small but potentially measurable deviations from Kerr~\cite{Vagnozzi:2020,Vagnozzi:2023,Kocherlakota:2021,Rahmatov:2026mdo}. The question is no longer whether alternatives exist. The question is whether we can tell any of them apart from Kerr with the data we have, or will have soon.

Wormholes sit near the top of that list of alternatives. A rotating wormhole can, with the right parameters, produce a shadow that looks almost identical to Kerr's when only the overall size is considered~\cite{Cunha:2018,Xavier:2023exm}. The differences show up in the details. Because there is no horizon, null rays can reach the throat and either scatter back out or pass through, changing the photon-ring structure and the splitting between prograde and retrograde critical impact parameters~\cite{Perlick:2015,Grenzebach:2014,Haroon:2019}. Whether these differences survive realistic accretion-flow modeling is an open question, but they are the right place to look~\cite{Kumar:2023wfp}. Meanwhile, a separate line of thinking has been developing on the thermodynamic side. The Bekenstein--Hawking area law ties horizon area to entropy~\cite{Bekenstein:1973,Hawking:1975}, and while the relation is robust, its microscopic origin remains unclear. Barrow recently proposed that quantum-gravitational effects might roughen the horizon into a fractal-like structure, replacing the area law with \(S \sim A^{1+\Delta/2}\), where \(0 \leq \Delta \leq 1\) measures how far the entropy departs from exact area scaling~\cite{Barrow:2020tzx}. The proposal has been used widely in cosmology~\cite{Saridakis:2020,Sheykhi:2021,Asghari:2022} and in black hole thermodynamics~\cite{Nojiri:2022,Capozziello:2025axh}. More relevant to our purposes, it fits naturally into an entropy--geometry correspondence in which the entropy functional determines the spacetime's effective matter content~\cite {Jusufi:2026jqb}. The practical advantage is that the resulting matter distributions are smooth rather than pointlike, which removes the usual singular behavior at small radii. Regular black holes offer a useful set of benchmarks for any such construction. The Bardeen metric was originally introduced as a toy model~\cite{Bardeen:1968}. It was later shown to arise naturally from nonlinear electrodynamics coupled to gravity~\cite{AyonBeato:1998}. Hayward's solution has a de Sitter-like core, a maximum mass, and an extremal limit at zero temperature~\cite {Hayward:2006}. Dymnikova's metric shares the vacuum-core feature~\cite{Dymnikova:1992}. These objects have been studied extensively---their thermodynamics, stability, and optical properties are reasonably well understood~\cite{Acevedo-Munoz:2025ueh,Wang:2023nwd}--- making them convenient reference points for comparing predictions.

The link between all of this and what a telescope sees runs through the photon sphere. For a static, spherically symmetric spacetime, the photon sphere sits where \(r f'(r) = 2 f(r)\), and the shadow radius follows from \(R_{\rm sh} = r_{\rm ph}/\sqrt{f(r_{\rm ph})}\)~\cite{Cunha:2018}. Rotation complicates the picture: the photon region becomes three-dimensional, and the shadow boundary has to be constructed from both prograde and retrograde orbits~\cite{Gralla:2020,Johnson:2020,Perlick:2015,Grenzebach:2014}. Wormholes modify the condition itself. For the geometry studied here, with $N=1$ on the equatorial plane, the static photon sphere coincides with the throat, $b(r_{\rm ph})=r_{\rm ph}$, while slow rotation shifts the prograde branch to $r_{\rm ph}^2\simeq 6J$. In principle, the throat can sit right at the photon sphere~\cite{Nedkova:2013msa,Ohgami:2015nra,Tangphati:2023mpk}. This is a real difference, not a cosmetic one, and it is the main reason wormhole and black hole shadows need not coincide even when their overall sizes do. This paper connects these threads in a single concrete model. We construct a rotating traversable wormhole whose matter content arises from a Barrow deformation of the Bekenstein--Hawking area law, with the deformation parameter \(n\) restricted to \(0 < n \leq 1\). The deformation smears the source over a finite region and replaces the undeformed solution's pointlike divergence with a regular profile. The energy density is negative everywhere and falls off as \(\rho_B \propto -r^{-(n+3)}\), and the shape function flares out smoothly at the throat, so traversability is built in rather than imposed by hand. We treat rotation in the slow-rotation regime, choosing the lapse and frame-dragging functions so both remain finite and positive at the throat. No horizon appears, and the geometry remains asymptotically flat.

The paper's structure is straightforward. Sect.~\ref{Sec:II} sets up the rotating wormhole geometry and the entropy-inspired matter smearing, ending with the full metric and a short discussion of regularity. Sect.~\ref{Sec:III} follows null geodesics in the equatorial plane, derives the photon-sphere condition, and works out the critical impact parameters and the resulting shadow. Sect.~\ref{Sec:IV} examines the energy conditions and the exotic-matter content, including the volume-integral quantifier. Sect.~\ref{Sec:V} compares the predicted shadow size, rotational asymmetry, and photon-ring separation with Kerr and with the Bardeen, Hayward, Ayon-Beato--Garcia, and Dymnikova regular black holes. Finally, Sect.~\ref{Sec:VI} summarizes the main results.

%-----------------------------------------------------------
\section{Rotating wormhole geometry and entropy-inspired matter smearing}
\label{Sec:II}
%-----------------------------------------------------------

This section establishes the geometric framework for a rotating traversable wormhole supported by a regular, entropy-inspired matter distribution. We begin with the stationary and axisymmetric wormhole metric, introduce the entropy--geometry correspondence that generates smeared matter profiles, and then construct the complete rotating solution.

%-----------------------------------------------------------
\subsection{Stationary and axisymmetric wormhole geometry}
%-----------------------------------------------------------

A rotating traversable wormhole is described by a spacetime that is both stationary and axisymmetric, admitting two commuting Killing vectors, \(X^\mu = \left(\frac{\partial}{\partial t}\right)^\mu\), \(Y^\mu = \left(\frac{\partial}{\partial \phi}\right)^\mu\), associated with time translations and rotations about the symmetry axis. Consequently, all metric functions are independent of the coordinates \(t\) and \(\phi\).

Under the simultaneous transformation \((t,\phi)\rightarrow(-t,-\phi)\), the line element takes the canonical form~\cite{Papapetrou:1966zz,Carter:1969zz}
\begin{eqnarray}
ds^2 = g_{tt}\,dt^2 + 2g_{t\phi}\,dt\,d\phi + g_{\phi\phi}\,d\phi^2
+ g_{rr}\,dr^2 + g_{\theta\theta}\,d\theta^2,~~~
\label{eq:canonicalmetric}
\end{eqnarray}
where the off-diagonal term \(g_{t\phi}\) encodes the frame-dragging effect due to rotation. Asymptotic flatness imposes the boundary conditions
\[
g_{tt}\rightarrow -1,\qquad g_{t\phi}\rightarrow 0,\qquad
g_{\phi\phi}\rightarrow r^2\sin^2\theta,\qquad (r\rightarrow\infty).
\]

Following Teo~\cite{Teo:1998dp}, the metric for a rotating traversable wormhole is written as
\begin{eqnarray}
ds^2 &=& -N^2(r,\theta)\,dt^2 + e^{\mu(r,\theta)}\,dr^2
+ r^2K^2(r,\theta)\Big[ d\theta^2 \nonumber\\
&& + \sin^2\theta \bigl(d\varphi - \omega(r,\theta)\,dt\bigr)^2 \Big],
\label{eq:rotmetric}
\end{eqnarray}
where \(N(r,\theta)\) is the redshift function, \(\mu(r,\theta)\) determines the spatial shape, \(K(r,\theta)\) modifies the angular sector, and \(\omega(r,\theta)\) represents the angular velocity of frame dragging. In the non-rotating limit, one recovers the standard Morris-Thorne wormhole with $N\to e^{\Phi(r)}$, $e^{\mu}\to(1-b(r)/r)^{-1}$, $K\to 1$, and $\omega\to 0$.

%-----------------------------------------------------------
\subsection{Entropy--geometry correspondence and matter smearing}
%-----------------------------------------------------------

To support the wormhole geometry with a regular matter source, we employ the entropy--geometry correspondence developed in Ref.~\cite{Anand:2025rjg}. This framework establishes a direct relation between black hole entropy and the spacetime's effective matter content, naturally smearing the source over a finite region.

Starting from the static, spherically symmetric line element
\begin{eqnarray}
ds^2 = -f(r)\,dt^2 + \frac{dr^2}{f(r)} + r^2 d\Omega^2,
\end{eqnarray}
with horizon at \(r=r_+\) satisfying \(f(r_+)=0\). The first law of thermodynamics \(dM = T_H\,dS\) and the Hawking temperature \(T_H = f'(r_+)/4\pi\) yield a relation between the metric function and the entropy. Extending this relation beyond the horizon gives
\begin{eqnarray}
f(r)=1-\frac{2M}{r}\,\frac{S_{\rm BH}}{S(r)}\,S'(r)\,\frac{r}{2},
\label{eq:entropy_metric}
\end{eqnarray}
which reduces to $f(r)=1-2M/r$ for the Bekenstein--Hawking entropy $S_{\rm BH}=4\pi r_+^2$. This shows that the geometry is entirely determined once the entropy functional \(S(r)\) is specified.

In Einstein gravity, the modified geometry in Eq.~\eqref{eq:entropy_metric} can be interpreted as being generated by an effective anisotropic fluid. The corresponding energy density is
\begin{eqnarray}
\rho(r)=-\,\frac{M}{8\pi r^{2}}\,
\frac{r\,S''(r)-S'(r)}{S'(r)^{2}},
\label{eq:rho_entropic_general}
\end{eqnarray}
so that $[\rho]=\text{length}^{-2}$ in units $G=c=1$. The radial pressure satisfies \(p_r(r) = -\rho(r)\), while the tangential pressure depends on higher derivatives of the entropy. The relation \(p_r = -\rho\) is a consequence of the specific metric ansatz, whereas the tangential pressure captures the anisotropic nature of the source.

This entropy-based construction naturally regularizes the matter distribution: point-like divergences are replaced by smeared profiles spread over a finite region. In the present analysis, we use the entropy-generated density profile \(\rho(r)\) as the wormhole source. Among various entropy deformations, we focus on the Barrow entropy model, which introduces a fractal deformation of the Bekenstein--Hawking area law via the parameter \(n\)~\cite{Barrow:2020tzx}. The corresponding density profile is
\begin{eqnarray}
\rho_B(r)=-\,\frac{n\,M}{2(n+2)\pi^{1+n/2}}\,
\frac{1}{r^{\,n+3}},\qquad 0<n\le 1,
\label{eq:rho_B2}
\end{eqnarray}
where the prefactor is fixed by dimensional analysis so that $[\rho_B]=\text{length}^{-2}$ %\fx{and so that $8\pi\rho_B r^2 = b'(r)$ holds exactly with the shape function defined below}. 
For positive \(M\) and \(n\), the density remains negative throughout spacetime. As \(n\rightarrow 0\), $\rho_B\to 0$ and the shape function approaches the constant $b(r)\to r_{\rm th}$; the Bekenstein--Hawking area law is recovered in the entropy, while the exotic source switches off. Thus, small \(n\) values track the transition from the entropy-deformed configuration to the undeformed case.

%============================================================================
\begin{table*}[!t]
\centering
\renewcommand{\arraystretch}{1.25}
\setlength{\tabcolsep}{4pt}
\small

{\footnotesize \caption{Barrow entropy-inspired shape function $b(r)$ and the single lapse function $N(r,\theta)$ used throughout this work, together with their impact on photon dynamics in slowly rotating wormholes.}
\label{tab:logfamily}

\begin{tabular}{@{}l l l@{}}
\toprule
{Model}
&
{Radial structure}
&
{Photon dynamics role}
\\
\midrule

{Barrow entropy}
&
$\displaystyle
b(r)=r_{\rm th}+\frac{4M}{(n+2)\pi^{n/2}}
\left(\frac{1}{r^n}-\frac{1}{r_{\rm th}^n}\right),
\qquad 0<n\leq1$
&
\parbox[t]{0.47\textwidth}{
Fractal deformation controls localization; smaller $n$ gives a sharper throat, while larger $n$ spreads the violation region.
}
\\[10pt]

{Lapse}
&
$\displaystyle
N(r,\theta)=\exp\!\left[\frac{\alpha J^{2}}{r^{2}}\cos^{2}\theta\right]$
&
\parbox[t]{0.47\textwidth}{
Positive and finite at the throat for all $\alpha$ and $J$;
$N\to 1$ as $r\to\infty$.
}
\\[10pt]

{Equatorial lapse}
&
$\displaystyle
N(r,\pi/2)=1$
&
\parbox[t]{0.47\textwidth}{
Simplifies photon dynamics; rotation enters through
$\displaystyle \omega(r)=\frac{2J}{r^3}$ only.
}
\\

\bottomrule
\end{tabular}}
\end{table*}
%============================================================================

%-----------------------------------------------------------
\subsection{From shape function to lapse: constructing the wormhole geometry}
%-----------------------------------------------------------
\label{sec:entropy_wormhole_geometry}

We build the wormhole geometry from the entropy-inspired matter distribution, but we let the redshift sector stand on its own rather than tying it to the shape function. For a static, spherically symmetric configuration, the metric takes the Morris--Thorne form
\begin{eqnarray}
ds^2 = -N^2(r)\,dt^2
+ \frac{dr^2}{1-b(r)/r}
+ r^2\,d\Omega^2,
\label{eq:static_wormhole_metric}
\end{eqnarray}
where $N(r)=e^{\Phi(r)}$ is the lapse (or redshift) function and $b(r)$ is the shape function. The throat sits at $r=r_{\rm th}$, defined by $b(r_{\rm th})=r_{\rm th}$. The flare-out condition then requires
\begin{eqnarray}
b'(r_{\rm th})<1.
\label{eq:flareout_condition}
\end{eqnarray}
For the wormhole to be traversable, the redshift function must also stay finite and nonzero at the throat,
\begin{eqnarray}
0<N(r_{\rm th})<\infty,
\label{eq:regular_lapse_condition}
\end{eqnarray}
which keeps the throat from turning into a Killing horizon.

The shape function follows from the entropy-inspired energy density through
\begin{eqnarray}
b(r) = 8\pi\int_0^r \rho_B(r')\,r'^2\,dr'.
\label{eq:ShapeFEntropy}
\end{eqnarray}
Because the Barrow profile~\eqref{eq:rho_B2} is a pure power law, the integral must be evaluated from the throat outward,
\begin{eqnarray}
b(r) = r_{\rm th} + 8\pi\int_{r_{\rm th}}^r \rho_B(r')\,r'^2\,dr',
\label{eq:ShapeFEntropy_corrected}
\end{eqnarray}
which yields
\begin{eqnarray}
b(r) = r_{\rm th}
+ \frac{4M}{(n+2)\pi^{n/2}}
\left(
\frac{1}{r^n}
-
\frac{1}{r_{\rm th}^n}
\right),
\label{eq:shape_barrow}
\end{eqnarray}
with $M$ the mass parameter, $r_{\rm th}$ the throat radius, and $n\in(0,1]$ the parameter that measures the entropy deformation. The construction is normalized so that $b(r_{\rm th})=r_{\rm th}$ exactly. Differentiating the shape function,
\begin{eqnarray}
b'(r) = -\frac{4Mn}{(n+2)\pi^{n/2}r^{n+1}},
\label{eq:bprime_barrow}
\end{eqnarray}
so that
\begin{eqnarray}
b'(r_{\rm th}) = -\frac{4Mn}{(n+2)\pi^{n/2}r_{\rm th}^{n+1}}.
\label{eq:flareout_barrow}
\end{eqnarray}
For $M>0$ and $n>0$ this is negative, and therefore
\begin{eqnarray}
b'(r_{\rm th})<1
\end{eqnarray}
holds automatically. It is handy to define the dimensionless combination
\begin{eqnarray}
\beta_n = \frac{4M}{(n+2)\pi^{n/2}r_{\rm th}^{n+1}},
\label{eq:beta_n}
\end{eqnarray}
which lets us write
\begin{eqnarray}
b'(r_{\rm th})=-n\beta_n,
\qquad
1-b'(r_{\rm th})=1+n\beta_n>0.
\label{eq:beta_flareout}
\end{eqnarray}
The entropy-inspired shape function thus flares out well at the throat.

The large-distance behavior also matters. As $r$ grows,
\begin{eqnarray}
b(r) \longrightarrow b_\infty = r_{\rm th}
- \frac{4M}{(n+2)\pi^{n/2}r_{\rm th}^{n}},
\qquad
r\rightarrow\infty,
\label{eq:b_asymptotic}
\end{eqnarray}
and so
\begin{eqnarray}
\frac{b(r)}{r}\longrightarrow0.
\label{eq:asymptotic_flatness_shape}
\end{eqnarray}
The geometry is asymptotically flat in the radial direction.

\begin{figure*}
\begin{center}
\includegraphics[width=14.2cm,height=7cm]{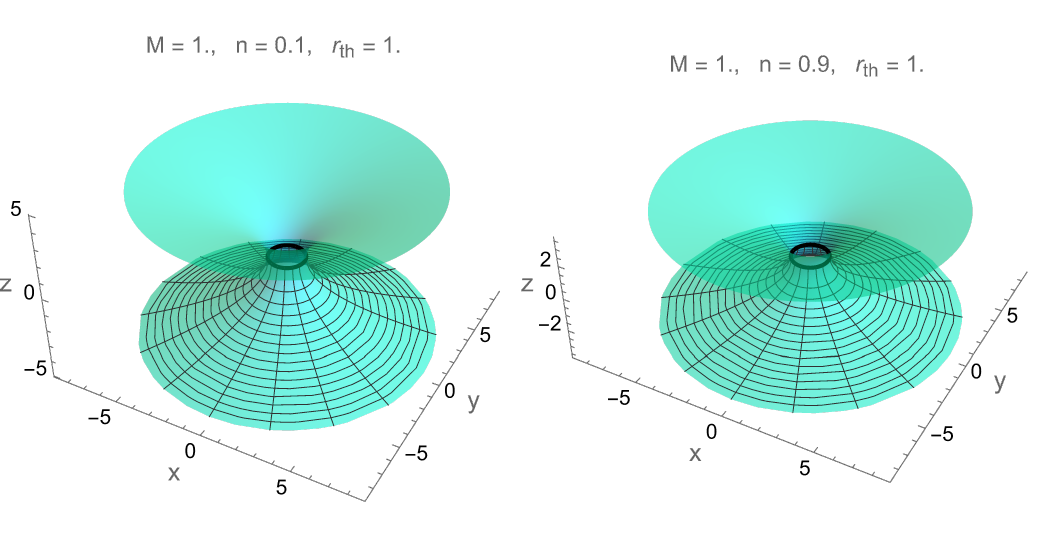}
\end{center}
\setlength{\abovecaptionskip}{0pt}
\setlength{\belowcaptionskip}{0pt}
\caption{{\footnotesize Embedding diagrams of the Barrow entropy-inspired wormhole geometry for \(M=1\), \(r_{\rm th}=1\), and two representative values of the Barrow parameter, \(n=0.1\) and \(n=0.9\). The surfaces smoothly flare out from the throat, \(r=r_{\rm th}\), illustrating the traversable wormhole geometry and the effect of increasing \(n\) on the spatial embedding.}}\label{Fig0}
\end{figure*}

The Barrow entropy tied to this deformation scales generically as
\begin{eqnarray}
S_B(r) = S_0\,\pi^{n/2}\,r^{\,2+n},
\label{eq:barrow_entropy}
\end{eqnarray}
which reproduces the area law $S_B\propto r^{2}$ as $n\to 0$. Taking $n\rightarrow0$ recovers the usual area law, so $n$ is the knob that controls how far the entropy departs from area scaling. The spatial embedding of the resulting geometry is shown in Fig.~\ref{Fig0}.

%----------------------------------------------------------------------
%----------------------------------------------------------------------

\subsection{Slowly rotating extension}
\label{sec:slow_rotation_geometry}

We now extend the static wormhole to slow rotation. To leading nontrivial order in the angular momentum,
\begin{eqnarray}
ds^2 ={}& -N^2(r,\theta)\,dt^2
+ \frac{dr^2}{1-b(r)/r}
\nonumber\\
&+ r^2K^2(r,\theta)
\left[
d\theta^2
+ \sin^2\theta
\left(d\varphi-\omega(r)\,dt\right)^2
\right],
\label{eq:slowrot_wormhole_metric}
\end{eqnarray}
where $b(r)$ is the shape function from Eq.~\eqref{eq:shape_barrow}, while the lapse and the angular deformation are taken as
\begin{eqnarray}
N(r,\theta) = \exp\left[\frac{\alpha J^2}{r^2}\cos^2\theta\right],
\label{eq:lapse_regular}
\end{eqnarray}
and
\begin{eqnarray}
K(r,\theta) = \exp\left[\frac{\beta J^2}{r^2}\cos^2\theta\right].
\label{eq:K_regular}
\end{eqnarray}
Here $J$ is the angular momentum, and $\alpha$ and $\beta$ are the only independent dimensionless parameters that control the leading rotational corrections to the lapse and the angular sector. Writing both in exponential form keeps them finite and positive everywhere in the physical region. Expanding for small $J$,
\begin{eqnarray}
N(r,\theta) = 1 +
\frac{\alpha J^2}{r^2}\cos^2\theta
+ \mathcal{O}(J^4),
\label{eq:lapse_expansion}
\end{eqnarray}
and
\begin{eqnarray}
K(r,\theta) = 1 +
\frac{\beta J^2}{r^2}\cos^2\theta
+ \mathcal{O}(J^4).
\label{eq:K_expansion}
\end{eqnarray}
Several features are worth noting. The lapse is positive for all $r\geq r_{\rm th}>0$ regardless of the sign of $\alpha$, it stays finite at the throat,
\begin{eqnarray}
N(r_{\rm th},\theta) = \exp\left[
\frac{\alpha J^2}{r_{\rm th}^2}\cos^2\theta
\right],
\label{eq:lapse_throat}
\end{eqnarray}
and it tends to unity at infinity,
\begin{eqnarray}
N(r,\theta) \longrightarrow 1,
\qquad
r\rightarrow\infty.
\label{eq:lapse_infinity}
\end{eqnarray}
The redshift function
\begin{eqnarray}
\Phi(r,\theta) = \ln N(r,\theta) = \frac{\alpha J^2}{r^2}\cos^2\theta
\label{eq:redshift_regular}
\end{eqnarray}
is then manifestly finite throughout the physical region, so no artificial horizon appears at the throat.

For the frame-dragging function, we keep the leading slow-rotation term,
\begin{eqnarray}
\omega(r) = \frac{2J}{r^3}.
\label{eq:omega_slowrotation}
\end{eqnarray}
This gives the ordering
\begin{eqnarray}
\omega(r)=\mathcal{O}(J),
~~
N-1=\mathcal{O}(J^2),
~~
K-1=\mathcal{O}(J^2),
\end{eqnarray}
which is exactly what slow rotation requires: frame dragging enters linearly in $J$, while the diagonal rotational deformations start at quadratic order.

The frame-dragging contribution to the metric goes as
\begin{eqnarray}
r^2K^2\sin^2\theta\,\omega^2(r) = \mathcal{O}(J^2),
\end{eqnarray}
so every rotational term we keep is consistently controlled up to $\mathcal{O}(J^2)$.

At the throat the radial coefficient still vanishes,
\begin{eqnarray}
1-\frac{b(r_{\rm th})}{r_{\rm th}}=0,
\end{eqnarray}
marking the minimal-area surface. This is a coordinate feature of the Morris--Thorne radial coordinate, not a horizon. The lapse there remains finite and positive,
\begin{eqnarray}
0<N(r_{\rm th},\theta)<\infty,
\end{eqnarray}
so the throat is not tied to a lapse zero.

Far away,
\begin{eqnarray}
N(r,\theta)\rightarrow1,
~~
K(r,\theta)\rightarrow1,
~~
\omega(r)\rightarrow0,
~~
\frac{b(r)}{r}\rightarrow0,
\end{eqnarray}
as $r\rightarrow\infty$, and the spacetime settles into an asymptotically flat, nonrotating geometry.

\begin{figure*}
\begin{center}
\includegraphics[width=17.5cm,height=8.5cm]{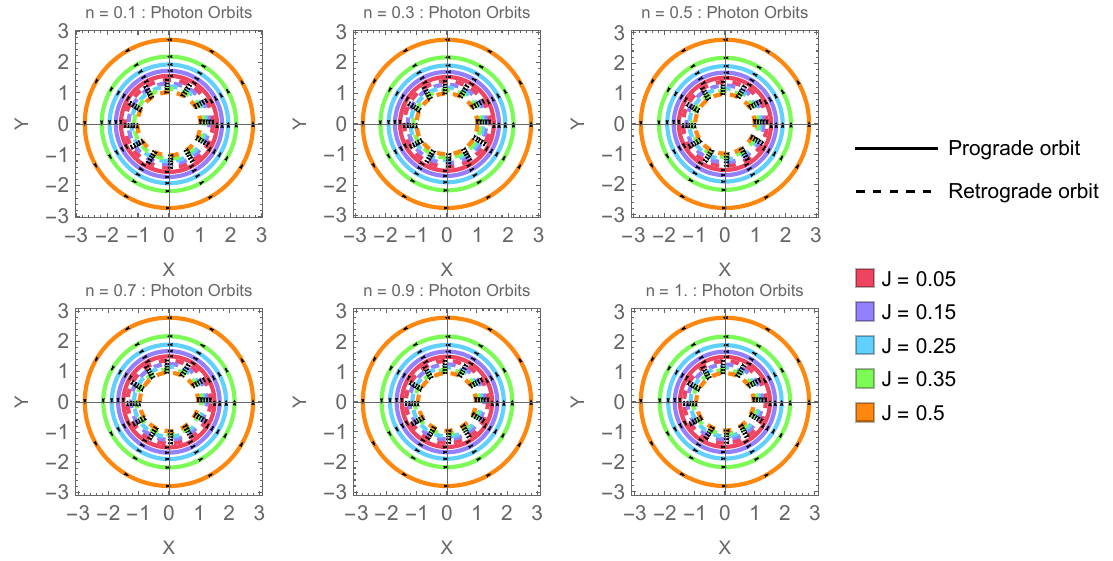}
\end{center}
\setlength{\abovecaptionskip}{0pt}
\setlength{\belowcaptionskip}{0pt}
\caption{{\footnotesize Photon-orbit structure of the rotating Barrow entropy-inspired wormhole for \(M=1\) and \(r_{\rm th}=1\), shown for \(n=0.1,\,0.3,\,0.5,\,0.7,\,0.9,\) and \(1.0\). The solid and dashed curves represent the prograde and retrograde photon orbits, respectively, while the color scale corresponds to \(J=0.05,\,0.15,\,0.25,\,0.35,\) and \(0.50\) (in units of \(M^2\)). Increasing the spin separates the two critical orbits, illustrating the rotational distortion of the photon region.}}\label{Fig1}
\end{figure*}

The construction cleanly separates the two ingredients. The entropy-inspired density fixes the shape function, and with it the throat and flare-out behavior, while the redshift function is chosen independently to keep the geometry regular and horizon-free. This split is especially convenient for the rotating case, since it lets us study the entropy deformation and the rotational corrections independently in the photon dynamics and in the shadow observables. The resulting photon-orbit structure, including the spin-dependent separation of prograde and retrograde branches, is illustrated in Fig.~\ref{Fig1}.

%-----------------------------------------------------
%---------------------------------------------------------------------

%-----------------------------------------------------------
\subsection{Regularity and energy conditions}
%-----------------------------------------------------------

Several regularity conditions must be imposed. The circumferential radius \(R = rK(r,\theta)\) must satisfy \(\partial R/\partial r > 0\) to maintain a regular embedding. The absence of horizons requires
\begin{eqnarray}
g_{t\varphi}^2 - g_{tt}g_{\varphi\varphi} = (NK\sin\theta)^2 > 0,
\end{eqnarray}
which holds as long as \(N>0\). On the symmetry axis \(\theta=0,\pi\), we require
\begin{eqnarray}
\partial_\theta N = \partial_\theta K = \partial_\theta \mu = 0,
\end{eqnarray}
to avoid angular singularities. In the local orthonormal frame, the NEC is examined using the null vector
\begin{eqnarray}
\kappa^\alpha = \left(\frac{1}{N},\ -e^{-\mu/2},\ 0,\ \frac{\omega}{N}\right).
\end{eqnarray}
The NEC requires \(R_{\alpha\beta}\kappa^\alpha\kappa^\beta \ge 0\). For the rotating wormhole metric, this quantity evaluates to~\cite{Teo:1998dp}
\begin{align}
R_{\alpha\beta}\kappa^\alpha\kappa^\beta &=
e^{-\mu}\mu_r\frac{(rK)_r}{rK}
- \frac{\omega_\theta^2\sin^2\theta}{2N^2}
- \frac{1}{4}\frac{\mu_\theta^2}{(rK)^2} \nonumber\\
&\quad - \frac{1}{2}\frac{(\mu_\theta\sin\theta)_\theta}{(rK)^2\sin\theta}
+ \frac{(N_\theta\sin\theta)_\theta}{(rK)^2N\sin\theta}.
\label{NECRotating}
\end{align}
Traversability demands that the NEC be violated in the vicinity of the throat, i.e.,
\begin{eqnarray}
R_{\alpha\beta}\kappa^\alpha\kappa^\beta < 0 \quad \text{near } r=r_{\rm th},
\end{eqnarray}
which is a fundamental requirement for wormhole geometries. The entropy-inspired matter distribution naturally accommodates this violation because of its negative energy density and anisotropic pressures. We treat rotation within the slow-rotation approximation, with the dimensionless parameter \(\chi = J/M^2 \ll 1\). This ensures that higher-order rotational corrections are negligible and that the leading-order frame-dragging function \(\omega(r)=2J/r^3\) provides an accurate description. We express all rotational quantities in dimensionless form using the natural mass scale \(M\), allowing consistent comparison between configurations. Thus, the entropy-inspired matter smearing, derived from the Barrow entropy model, yields a fully regular, rotating traversable wormhole geometry. The construction proceeds logically from the entropy functional to the density profile, then to the shape function, and finally to the full metric with a well-defined lapse and frame-dragging sector.

\section{Photon orbits near a slowly rotating entropy-generated wormhole}\label{Sec:III}

The wormhole geometry constructed in Sec.~\ref{Sec:II} emerges from a smeared entropy distribution rather than a pointlike source. Because the underlying matter is spread out, the spacetime is regular and horizonless. Adding a small amount of rotation breaks spherical symmetry in a controlled way and introduces frame dragging. In this section we study how light moves through such a spacetime, focusing on the equatorial plane where the effects of rotation are simplest to isolate.

\subsection{Frame dragging in a slowly rotating wormhole}
Rotation is introduced perturbatively by assuming that the wormhole rotates slowly. Let $J$ denote the total angular momentum and define the dimensionless spin parameter $\chi = J/M^2 \ll 1$. Keeping only first-order terms in $\chi$, the metric develops an off-diagonal frame-dragging contribution governed by
$\omega(r) = 2J/r^3$ [Eq.~\eqref{eq:omega_slowrotation}], while the lapse function receives a small rotational correction, $N(r,\theta) = 1 + \frac{\alpha J^2}{r^2}\cos^2\theta + \mathcal{O}(J^4)$
[Eq.~\eqref{eq:lapse_expansion}]. Here $\alpha$ is the single dimensionless parameter controlling how strongly rotation deforms the redshift factor. On the equatorial plane $\theta=\pi/2$ we have $N=1$ to leading order, so the metric simplifies considerably.

%----------------------------------------------------------------------

\begin{figure*}
\begin{center}
\includegraphics[width=17.5cm,height=8.5cm]{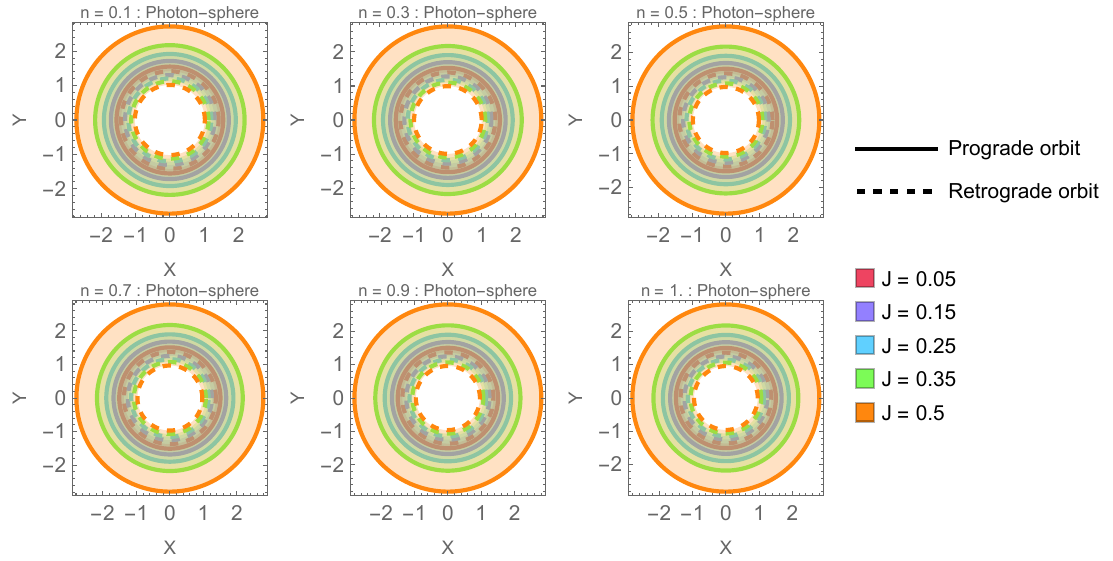}
\end{center}
\setlength{\abovecaptionskip}{0pt}
\setlength{\belowcaptionskip}{0pt}
\caption{{\footnotesize Photon-sphere structure of the rotating Barrow entropy-inspired wormhole for \(M=1\) and \(r_{\rm th}=1\), with \(n=0.1,\,0.3,\,0.5,\,0.7,\,0.9,\) and \(1.0\). The solid and dashed curves denote the prograde and retrograde photon orbits, respectively, while the shaded bands between them highlight the spin-dependent separation of the two critical impact parameters for \(J=0.05,\,0.15,\,0.25,\,0.35,\) and \(0.50\) (in units of \(M^2\)). The widening of these shaded regions with increasing \(J\) illustrates the growing rotational asymmetry of the photon-sphere structure.}}\label{Fig2}
\end{figure*}
%----------------------------------------------------------------------

\subsection{How fast does the local frame precess?}

A gyroscope at rest relative to the distant stars will nevertheless precess because the wormhole drags inertial frames. The Lense-Thirring (LT) precession frequency in the equatorial plane is
\begin{eqnarray}
\Omega_{\rm LT}(r)= -\frac12\frac{d\omega}{dr}= \frac{3J}{r^4}.
\label{eq:lt}
\end{eqnarray}
At the throat this reaches its maximum,
\begin{eqnarray}
\Omega_{\rm LT}(r_{\rm th}) = \frac{3J}{r_{\rm th}^4},
~\text{or in dimensionless form}~\nonumber\\
\lambda_{\rm LT} \equiv \Omega_{\rm LT}(r_{\rm th})\, r_{\rm th}^2
= \frac{3J}{r_{\rm th}^2}.
\label{eq:lt_dim}
\end{eqnarray}
Choosing representative numbers that satisfy the flare-out condition, say $r_{\rm th}=1.7M$ and $J=0.06M^2$, gives $\lambda_{\rm LT}\approx0.062$. Such small values justify the slow-rotation expansion throughout.

One might ask whether an ergoregion could appear. That would require $g_{tt}>0$, i.e. $4J^2/r^4 > N^2$. With $N\approx1$ on the equator and our tiny $J$, this never happens. Hence the wormhole remains traversable and free of horizons, exactly as desired.

\subsection{Light rays in the equatorial plane}

Because the problem is stationary and axisymmetric, a photon moving in the equatorial plane ($\theta=\pi/2$, $\dot\theta=0$) conserves energy and angular momentum. Write the metric as
\begin{eqnarray}
ds^2 = -dt^2 + \frac{dr^2}{1-b(r)/r}
+ r^2\bigl(d\varphi - \omega(r)dt\bigr)^2,
\label{eq:eq_metric}
\end{eqnarray}
where we have set $N=1$ as discussed. Let $\tilde E$ and $\tilde L$ be the conserved quantities per unit mass and define the impact parameter $\mathscr{B} = \tilde L/\tilde E$. Eliminating $\dot t$ and $\dot\varphi$ from the geodesic equations leads to the radial equation
\begin{eqnarray}
\left(\frac{dr}{d\lambda}\right)^2
= \left(1-\frac{b(r)}{r}\right)\tilde E^2 \, F(r),
\label{eq:radial_eq}
\end{eqnarray}
where
\begin{eqnarray}
F(r) = \bigl(1 + \mathscr{B} \,\omega(r)\bigr)^2
- \frac{\mathscr{B}^{2}}{r^{2}}.
\label{eq:F_def}
\end{eqnarray}

%---------------------------------------------------------------------
\begin{figure*}
\begin{center}
\includegraphics[width=17.5cm,height=9cm]{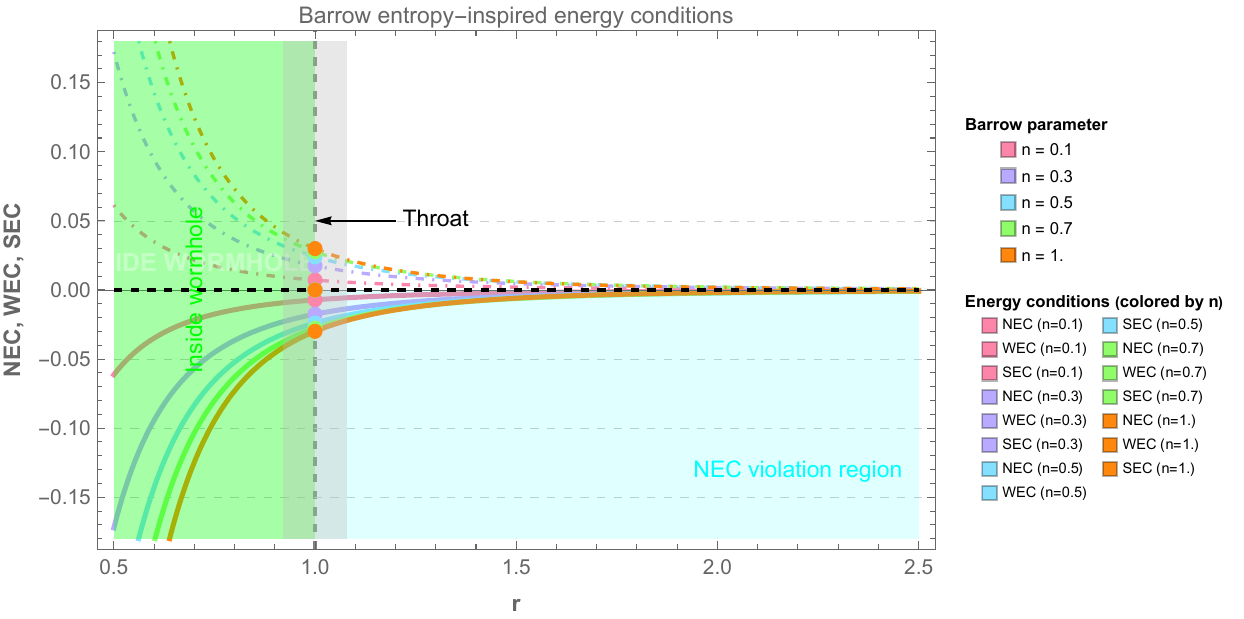}
\end{center}
\setlength{\abovecaptionskip}{0pt}
\setlength{\belowcaptionskip}{0pt}
\caption{{\footnotesize Energy condition profiles of the Barrow entropy-inspired wormhole for \(M=1\) and \(r_{\rm th}=1\), with \(n=0.1,\,0.3,\,0.5,\,0.7,\) and \(1.0\). The NEC, WEC, and SEC are shown by dashed, solid, and dot-dashed curves, respectively, with colors indicating the different values of \(n\). The green shaded region marks the interior of the wormhole (\(r\leq r_{\rm th}\)), while the cyan region highlights the exterior domain where the NEC is violated. The throat at \(r=r_{\rm th}=1\) is indicated by the vertical band and intersection points, illustrating the localization and \(n\)-dependence of the energy condition violations.}}\label{fig:energyconditions}
\end{figure*}
%----------------------------------------------------------------------

The denominator $1-r^2\omega^2$ that ordinarily appears has been set to one because $r^2\omega^2 = 4J^2/r^4$ is negligible at the order we work. Similarly, the azimuthal motion obeys
\begin{eqnarray}
\frac{d\varphi}{dr}
= \frac{\mathscr{B}/r^2 + \omega(r)}
{\sqrt{1-b(r)/r}\,\sqrt{F(r)}} \equiv G(r).
\label{eq:phi_eq}
\end{eqnarray}
For a photon that comes in from infinity, reaches a minimum radius $r_{\rm min}$, and then goes back out, the total deflection angle is
\begin{eqnarray}
\hat\alpha = -\pi + 2\int_{r_{\rm min}}^{\infty} G(r)\,dr.
\label{eq:deflection}
\end{eqnarray}
The sign of $\mathscr{B}$ distinguishes prograde ($\mathscr{B} >0$) from retrograde ($\mathscr{B} <0$) orbits. Because $\omega(r)$ is positive, the two cases yield different $F(r)$ and therefore different deflections, a clear signature of rotation.

\subsection{Where can a photon circle indefinitely?}

Circular photon orbits satisfy $\dot r =0$ and $\ddot r =0$. From
$\dot r=0$ we obtain $F(r)=0$, i.e.
\begin{eqnarray}
\bigl(1 + \mathscr{B} \,\omega(r)\bigr)^2 = \frac{\mathscr{B}^{2}}{r^{2}},
\label{eq:Fzero}
\end{eqnarray}
which gives two families of impact parameters,
\begin{eqnarray}
\mathscr{B}_{+}(r) = \frac{r}{1 - r\,\omega(r)},\qquad
\mathscr{B}_{-}(r) = -\,\frac{r}{1 + r\,\omega(r)}.
\label{eq:p_pm}
\end{eqnarray}
The extremum condition $\frac{d}{dr}\bigl[(1-b/r)F\bigr]=0$ at $F=0$ requires $F'(r_{\rm ph})=0$, provided the prefactor $1-b/r$ does not vanish at the photon sphere. Substituting the explicit form of $F$ and using $F(r_{\rm ph})=0$, we obtain
\begin{eqnarray}
F'(r_{\rm ph}) = 2\mathscr{B}^2\left(\pm\frac{\omega'(r_{\rm ph})}{r_{\rm ph}} + \frac{1}{r_{\rm ph}^3}\right) = 0,
\label{eq:ph_cond_general}
\end{eqnarray}
where the upper (lower) sign corresponds to $\mathscr{B}_+$ ($\mathscr{B}_-$). For the frame-dragging function $\omega(r)=2J/r^3$ we have $\omega'(r) = -6J/r^4$, and Eq.~\eqref{eq:ph_cond_general} becomes
\begin{eqnarray}
2\mathscr{B}^2\left(\mp\frac{6J}{r_{\rm ph}^5} + \frac{1}{r_{\rm ph}^3}\right) = 0.
\label{eq:ph_cond_J}
\end{eqnarray}
For the prograde branch ($\mathscr{B}_+$, upper sign), the bracket vanishes when
\begin{eqnarray}
r_{\rm ph}^2 = 6J \quad (\text{prograde}),
\label{eq:ph_prograde}
\end{eqnarray}
so that the prograde photon sphere sits at $r_{\rm ph} = \sqrt{6J}$. For the retrograde branch ($\mathscr{B}_-$), the bracket never vanishes for $r_{\rm ph}>0$ at leading order in $J$, so the retrograde circular orbit is pushed to the throat in the slow-rotation expansion.

In the static limit $J\to0$, the extremum condition degenerates, and the photon sphere is instead determined by the vanishing of the prefactor $1-b(r)/r$. In that case
\begin{eqnarray}
b(r_{\rm ph}) = r_{\rm ph}\quad (J=0).
\label{eq:ph_cond_static}
\end{eqnarray}
So the static photon sphere coincides with the throat. This replaces the black-hole relation $3b-rb'=2r$ and reflects the fact that, in the absence of rotation, null geodesics on the equatorial plane of a Morris--Thorne wormhole with $N=1$ cannot circulate at any radius other than the throat itself.

For the Barrow shape function~\eqref{eq:shape_barrow}, Eq.~\eqref{eq:ph_cond_static} is satisfied identically at $r=r_{\rm th}$, since $b(r_{\rm th})=r_{\rm th}$ by construction. The leading rotational correction then shifts the prograde photon sphere to $r_{\rm ph}\simeq\sqrt{6J}$ and the critical impact parameters follow from evaluating Eq.~\eqref{eq:p_pm} at this radius.

Evaluating $\mathscr{B}_{\pm}(r)$ at $r=r_{\rm ph}$ gives the critical
impact parameters that define the shadow edge:
\begin{eqnarray}
\mathscr{B}_{+} = \frac{r_{\rm ph}}{1 - 2J/r_{\rm ph}^{2}},\qquad
\mathscr{B}_{-} = -\,\frac{r_{\rm ph}}{1 + 2J/r_{\rm ph}^{2}}.
\label{eq:p_crit}
\end{eqnarray}
Expanding for small $J$,
\begin{eqnarray}
\mathscr{B}_{+} \approx r_{\rm ph} + \frac{2J}{r_{\rm ph}},\qquad
\mathscr{B}_{-} \approx -\,r_{\rm ph} + \frac{2J}{r_{\rm ph}}.
\label{eq:p_crit_exp}
\end{eqnarray}

%---------------------------------------------------------------------
\begin{figure*}
\begin{center}
\includegraphics[width=17cm,height=8cm]{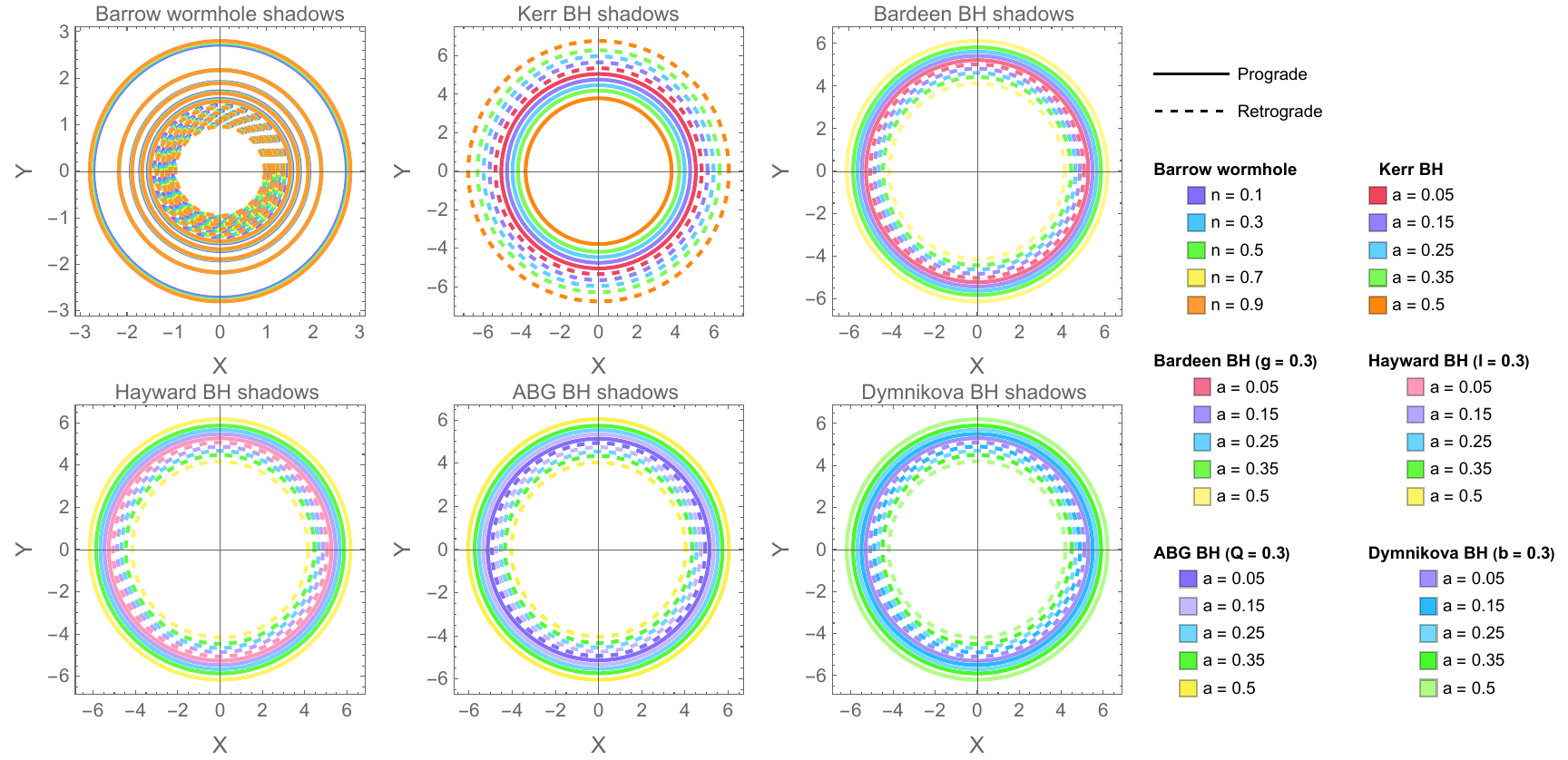}
\end{center}
\setlength{\abovecaptionskip}{0pt}
\setlength{\belowcaptionskip}{0pt}
\caption{{\footnotesize Comparison of the shadow structures of the rotating Barrow entropy-inspired wormhole and regular black hole models for \(M=1\) and \(r_{\rm th}=1\). The Barrow wormhole is shown for \(n=0.1,\,0.3,\,0.5,\,0.7,\) and \(0.9\), while the Kerr, Bardeen, Hayward, ABG, and Dymnikova black holes are considered for \(a/M=0.05,\,0.15,\,0.25,\,0.35,\) and \(0.50\), with \(g=\ell=Q=d=0.3M\), respectively. Solid and dashed curves denote the prograde and retrograde critical orbits. The comparison highlights how the Barrow parameter and rotation modify the shadow size and left-right asymmetry relative to regular black hole geometries.}}\label{fig:shadow_comparison}
\end{figure*}
%----------------------------------------------------------------------

The full splitting is therefore
$\Delta \mathscr{B} \equiv \mathscr{B}_{+} - \mathscr{B}_{-}
\approx 2r_{\rm ph}$, while the rotation-induced asymmetry is
$(\mathscr{B}_{+}+\mathscr{B}_{-})/2 \approx 2J/r_{\rm ph}$. Because the prograde photon sphere satisfies $r_{\rm ph}^2 = 6J$, the asymmetry scales as $2J/r_{\rm ph} = r_{\rm ph}/3$, which grows with spin as $\sqrt{J}$. The corresponding photon-sphere structure and the widening of the prograde--retrograde separation with spin are shown in Fig.~\ref{Fig2}.

\subsection{Does rotation shift the photon-sphere radius?}

A consistent expansion in $J$ shows that the leading shift
$\delta r$ of the photon-sphere radius is of order $J^{2}$ on the
equatorial plane. The dominant rotational effect is therefore the
asymmetry between $\mathscr{B}_{+}$ and $\mathscr{B}_{-}$ rather than a
displacement of $r_{\rm ph}$ itself. We work to leading order in $J$ and
do not quote an explicit $\delta r$ formula here.

\subsection{The shadow as a probe of entropy and rotation}

For a distant observer, the boundary of the wormhole shadow is traced by
the critical impact parameters $\mathscr{B}_{\pm}$. Therefore:
\begin{itemize}
\item In the nonrotating limit $J=0$, the shadow is a perfect circle of
radius $r_{\rm th}$.
\item For $J>0$, the circle becomes asymmetric: the prograde side
($\mathscr{B}_+$) is pushed outward, the retrograde side
($\mathscr{B}_-$) is pulled inward.
\item The fractional asymmetry
$\Delta\mathscr{B}/\mathscr{B}_{\rm avg} \approx 2J/r_{\rm ph}^{2}$
equals $1/3$ in the slow-rotation limit, since $r_{\rm ph}^2 = 6J$.
\end{itemize}

If the observer is not exactly in the equatorial plane, the lapse
function $N(r,\theta)$ from Eq.~\eqref{eq:lapse_regular} comes into play. For
an inclination angle $\theta_{\rm obs}$ (measured from the rotation
axis), the effective impact parameters become
\begin{eqnarray}
\mathscr{B}_{\pm}(\theta_{\rm obs})
= \frac{r_{\rm ph}}{N(r_{\rm ph},\theta_{\rm obs})}
\left[1 \pm \frac{2J}{r_{\rm ph}^{2}\,N(r_{\rm ph},\theta_{\rm obs})}\right].
\label{eq:p_inclined}
\end{eqnarray}
In the equatorial limit $N\to1$ this reduces correctly to Eq.~\eqref{eq:p_crit}. Because $N>1$ away from the equator, the shadow shrinks slightly and the asymmetry changes. In principle, comparing shadows from different viewing angles, or from different wormholes with random orientations, could constrain the rotational deformation parameter $\alpha$.

%============================================================================
\begin{table*}[!t]
\centering
\renewcommand{\arraystretch}{1.15}
\setlength{\tabcolsep}{5pt}
\small

\caption{{\footnotesize Observable quantities and their dependence on the underlying wormhole parameters.}}
\label{tab:summary_final}
{\footnotesize
\begin{tabular}{lcc}
\toprule
{Observable}
&
{Expression}
&
{What it encodes}
\\
\midrule

Photon-sphere radius (prograde)
&
$r_{\rm ph}^2 = 6J$
&
Spin $J$; independent of $n$ at leading order
\\[4pt]

Photon-sphere radius (static)
&
$b(r_{\rm ph}) = r_{\rm ph}$
&
Entropy exponent $n$ via throat
\\[4pt]

Shadow splitting
&
{$\displaystyle \Delta\mathscr{B}\approx 2r_{\rm ph}$,
asymmetry $\displaystyle \frac{2J}{r_{\rm ph}}$}
&
{Spin $J$ and $n$ via $r_{\rm ph}$}
\\[4pt]

Inclination dependence
&
{$\displaystyle
\mathscr{B}_{\pm}(\theta_{\rm obs})
\propto
\frac{1}{N(r_{\rm ph},\theta_{\rm obs})}$}
&
{Rotational parameter $\alpha$}
\\

\bottomrule
\end{tabular}}
\end{table*}
%============================================================================

%-----------------------------------------------------------
\section{Energy conditions and exotic matter content}\label{Sec:IV}
%-----------------------------------------------------------

The matter supporting a traversable wormhole is subject to strong constraints from the classical energy conditions. In particular, the throat's flare-out generally requires a departure from ordinary matter behavior, making analysis of the stress--energy tensor an important part of the construction. Here we examine these conditions for the entropy-inspired source introduced in Sec.~\ref{Sec:II}. Although the Barrow prescription modifies the matter distribution and provides a regular geometric description, it does not, by itself, remove the source's exotic character.

We describe the effective matter content by an aniso-tropic stress--energy tensor of the form
\begin{eqnarray}
T^\mu_{\;\nu}
=
{\rm diag}
\left(
-\rho,
p_r,
p_t,
p_t
\right),
\end{eqnarray}
where \(\rho\) denotes the energy density, while \(p_r\) and \(p_t\) are the radial and tangential pressures, respectively. In this representation, the NEC requires
\begin{eqnarray}
\rho+p_r \ge 0,
\qquad
\rho+p_t \ge 0,
\end{eqnarray}
whereas the WEC supplements these inequalities with
\begin{eqnarray}
\rho\geq0.
\end{eqnarray}
The SEC requires the NEC conditions together with
\begin{eqnarray}
\rho+p_r+2p_t\geq0.
\end{eqnarray}

For the rotating geometry considered here, the matter variables follow directly from the Einstein equations once we insert the Barrow-inspired shape function in Eq.~\eqref{eq:shape_barrow}. The resulting energy density takes the form
$\rho_B(r) = -\frac{n M}{2(n+2)\pi^{1+n/2}}\,
\frac{1}{r^{\,n+3}} < 0$ as given in Eq.~\eqref{eq:rho_B2}. It is therefore negative throughout the parameter range considered here. Consequently, the WEC is violated regardless of the detailed behavior of the pressure terms. This negative density is not an incidental feature of the solution; it reflects the exotic effective source required to sustain the wormhole geometry. The radial and tangential pressures also indicate where the energy condition violations are concentrated. Their combinations with the density, namely \(\rho+p_r\), \(\rho+p_t\), and \(\rho+p_r+2p_t\), are displayed in Fig.~\ref{fig:energyconditions} for representative values of the Barrow parameter \(n\). The most pronounced deviations occur close to the throat, where the geometric constraints associated with the flare-out condition are strongest. Away from the throat, the magnitude of the deviations decreases and the matter contribution becomes progressively weaker as the spacetime approaches its asymptotic regime. The parameter \(n\) controls how rapidly the entropy-induced source changes with radius. For smaller \(n\), the non-standard contribution is more strongly concentrated around the throat, resulting in a relatively narrow region where the NEC combinations are negative. Increasing \(n\) changes this radial profile and extends the region over which the corresponding energy conditions are violated. Thus, the Barrow deformation does not simply alter the magnitude of the exotic contribution; it also changes its radial distribution. The SEC follows the same general pattern. Its violation is most evident near the throat, where the pressures and energy density combine to sustain the nontrivial wormhole geometry. At larger radii, the relevant combinations decrease in magnitude as the geometry approaches the asymptotically flat region.

Overall, the analysis shows that the source's exotic character is unavoidable in the present Barrow-inspired construction, as already signaled by the negative energy density. At the same time, the pressure combinations reveal that the strongest departures from the classical energy conditions are associated with the throat region. The Barrow parameter therefore provides a direct way of modifying the radial distribution of the effective exotic matter while retaining the underlying traversable wormhole structure.

%============================================================================
\begin{table*}[!t]
\centering
\caption{{\footnotesize Comparison of entropy-inspired wormhole models with other regular and smeared-geometry constructions.}}
\label{tab:comparison}

\small
{\footnotesize
\setlength{\tabcolsep}{4pt}
\renewcommand{\arraystretch}{1.25}

\begin{tabular}{@{}l l l l@{}}
\toprule
{Feature}
&
{Regular BHs}
&
{Entropy-inspired models}
&
{This work}
\\
\midrule

{Matter source}
&
\parbox[t]{0.20\textwidth}{Regularized core\\
(Hayward/Bardeen)}
&
\parbox[t]{0.27\textwidth}{Entropy--geometry correspondence\\
\cite{Anand:2025rjg}}
&
\parbox[t]{0.27\textwidth}{Barrow entropy deformation\\
\cite{Barrow:2020tzx}}
\\[8pt]

{Energy density}
&
\parbox[t]{0.20\textwidth}{Nonlinear electrodynamics}
&
\parbox[t]{0.27\textwidth}{
Determined by $S(r)$ through
$\displaystyle
\rho=-\frac{M}{8\pi r^{2}}\frac{rS''-S'}{S'^{2}}$
}
&
\parbox[t]{0.27\textwidth}{
{$\displaystyle
\rho_B(r)=-\frac{n M}{2(n+2)\pi^{1+n/2}}
\frac{1}{r^{\,n+3}}$,\quad $0<n\le 1$}
}
\\[12pt]

{Radial pressure $p_r$}
&
\parbox[t]{0.20\textwidth}{Model-dependent anisotropy}
&
\parbox[t]{0.27\textwidth}{
$p_r=-\rho$ from the metric ansatz
}
&
\parbox[t]{0.27\textwidth}{
$p_r=-\rho$, consistent with the
entropy--geometry relation
}
\\[8pt]

{Tangential sector}
&
\parbox[t]{0.20\textwidth}{Standard anisotropic stress}
&
\parbox[t]{0.27\textwidth}{
Fixed by derivatives of $S(r)$
}
&
\parbox[t]{0.27\textwidth}{
Determined by $b(r)$ and $N(r)$
through the Einstein equations
}
\\[8pt]

{Geometry}
&
\parbox[t]{0.20\textwidth}{
Regular horizonless core
\cite{Hayward:2006,Bardeen:1968}
}
&
\parbox[t]{0.27\textwidth}{
Horizonless, entropy-generated metric
}
&
\parbox[t]{0.27\textwidth}{
Traversable wormhole with
$\displaystyle
b(r)=r_{\rm th}+\frac{4M}{(n+2)\pi^{n/2}}
\left(\frac{1}{r^n}-\frac{1}{r_{\rm th}^n}\right)$
}
\\

\bottomrule
\end{tabular}}
\end{table*}
%===========================================================================

\subsection{Volume-integral quantifier}\label{subsec:volume_integral}

The NEC violation near the throat provides a local diagnostic of the matter supporting the wormhole. It is also useful, however, to quantify this violation over the entire region outside the throat. For this purpose, we employ the volume-integral quantifier introduced by Visser, Kar, and Dadhich~\cite{Visser:2003yf}. In the present notation, it is defined as
\begin{eqnarray}
I_V=8\pi\int_{r_{\rm th}}^{\infty}
r^2\left(\rho+p_r\right)\,dr .
\end{eqnarray}
Thus, rather than characterizing the NEC at individual radii, \(I_V\) provides an integrated measure of the effective exotic matter associated with the wormhole. To evaluate this quantity, we first express the combination \(\rho+p_r\) in terms of the metric functions. For a static, spherically symmetric wormhole described by the shape function \(b(r)\) and lapse function \(N(r)\), the radial NEC combination can be written as
\begin{eqnarray}
\rho+p_r=
\frac{1}{8\pi r^2}
\left[b'(r)-\frac{b(r)}{r}\right]
+\frac{N'(r)}{8\pi r^2}
\left(1-\frac{b(r)}{r}\right).~~
\end{eqnarray}
Substitution into the volume integral then gives
\begin{small}
\begin{eqnarray}
I_V=
\int_{r_{\rm th}}^{\infty}
\left[b'(r)-\frac{b(r)}{r}\right]dr
+
\int_{r_{\rm th}}^{\infty}
r\left(1-\frac{b(r)}{r}\right)
\frac{N'(r)}{N(r)}\,dr .
\label{eq:IV_general}~~~
\end{eqnarray}
\end{small}

For the Barrow entropy-inspired construction considered here, the two metric functions entering this expression are given by Eqs.~\eqref{eq:shape_barrow} and~\eqref{eq:lapse_regular}. Consequently, once the model parameters are specified, the integrated quantity \(I_V\) can be evaluated directly. Its dependence on the throat scale and the Barrow deformation parameter \(n\) provides a global characterization of the exotic matter content, complementing the pointwise energy condition analysis presented above.

%============================================================================
\begin{table*}[!t]
\centering
\renewcommand{\arraystretch}{0.95}
\setlength{\tabcolsep}{2.5pt}
{\footnotesize
\caption{Compact comparison of energy-condition-violation features in different wormhole models.}
\label{tab:comparison_NEC2}

\begin{tabular}{lccc}
\toprule
Model & NEC violation & Localization & \(I_V\) \\
\midrule

Phantom scalar wormholes
& Strong, global
& Spacetime-wide
& Finite, large \\

Thin-shell wormholes
& Surface-layer violation
& At junction shell
& Finite, tension-dependent \\

Morris--Thorne class
& Model-dependent
& Near throat
& Adjustable \\

This work (Barrow entropy)
& \makecell[l]{\(p_r=-\rho<0\) everywhere;\\
mild violation strength}
& \makecell[l]{Throat-localized via \(n\);\\
\(\Delta r_{\rm ex}/r_{\rm th}\sim0.1{-}0.4\)}
& \makecell[l]{$I_V$ finite for\\
$r_{\rm th}^{n+1}=4M/[(n+2)\pi^{n/2}]$}
\\
\bottomrule
\end{tabular}
}
\end{table*}
%============================================================================

\subsubsection{Zero-tidal-force configuration}

For a zero-tidal-force configuration, the redshift function is constant, so that \(N'(r)=0\). The contribution involving the derivative of the lapse therefore drops out of the volume-integral quantifier, leaving
\begin{eqnarray}
I_V=
\int_{r_{\rm th}}^{\infty}
\left[b'(r)-\frac{b(r)}{r}\right]dr .
\label{eq:IV_zrf}
\end{eqnarray}
Using the entropy-inspired shape function in Eq.~\eqref{eq:shape_barrow} gives
\begin{eqnarray}
b'(r)=
-\frac{4Mn}{(n+2)\pi^{n/2}\,r^{\,n+1}},
\label{eq:bp_barrow}
\end{eqnarray}
and hence
\begin{eqnarray}
b'(r)-\frac{b(r)}{r}
=
-\frac{r_{\rm th}}{r}
-\frac{4M}{(n+2)\pi^{n/2}\,r}
\left[
\frac{n+1}{r^{n}}
-
\frac{1}{r_{\rm th}^n}
\right].
\label{eq:integrand_zrf}
\end{eqnarray}
This expression also shows that the asymptotic behavior of the volume integral must be handled with care. In particular, because \(b(r)\to r_{\rm th}\) as \(r\to\infty\) for the shape function, the logarithmic contribution is modified accordingly. One can obtain a finite value for the exotic-matter measure by introducing a finite radial cutoff \(R\) or by adopting an asymptotically decaying shape function.

For a finite integration domain \(r_{\rm th}\leq r\leq R\), one obtains
\begin{eqnarray}
I_V(R)
&=&
\frac{4M}{(n+2)\pi^{n/2}}
\left(\frac{1}{R^n}-\frac{1}{r_{\rm th}^n}\right)
-r_{\rm th}\ln\!\left(\frac{R}{r_{\rm th}}\right)
\nonumber\\&&-\frac{4M}{(n+2)\pi^{n/2}}
\int_{r_{\rm th}}^{R}\frac{1}{r}\left(\frac{1}{r^n}-\frac{1}{r_{\rm th}^n}\right)dr,
\label{eq:IV_R_zrf}
\end{eqnarray}
where the precise interpretation of \(r_{\rm th}\) should be tied to the actual throat condition \(b(r_{\rm th})=r_{\rm th}\). This form makes explicit that the integrated NEC contribution depends on both the deformation parameter and the radial scale over which the exotic matter is counted.

Evaluating the remaining integral explicitly,
\begin{eqnarray}
\int_{r_{\rm th}}^{R}\frac{1}{r}
\left(\frac{1}{r^n}-\frac{1}{r_{\rm th}^n}\right)dr
&=&
-\frac{1}{n}\left(\frac{1}{R^n}-\frac{1}{r_{\rm th}^n}\right)
-\frac{1}{r_{\rm th}^n}\nonumber\\&&\times\ln\!\left(\frac{R}{r_{\rm th}}\right),
\end{eqnarray}
so that Eq.~\eqref{eq:IV_R_zrf} becomes
\begin{eqnarray}
I_V(R)
&=&
\frac{4M(n+1)}{n(n+2)\pi^{n/2}}
\left(\frac{1}{R^n}-\frac{1}{r_{\rm th}^n}\right)
\nonumber\\&&+
\left[
\frac{4M}{(n+2)\pi^{n/2}\,r_{\rm th}^n}
-r_{\rm th}
\right]
\ln\!\left(\frac{R}{r_{\rm th}}\right).
\label{eq:IV_R_final_zrf}
\end{eqnarray}
For convergence as \(R\to\infty\), the coefficient of the logarithmic term must vanish, which requires
\begin{eqnarray}
r_{\rm th}^{n+1}=\frac{4M}{(n+2)\pi^{n/2}}.
\label{eq:finite_condition}
\end{eqnarray}
If this condition holds, the logarithmic divergence cancels and \(I_V\) remains finite. Otherwise, we need a finite cutoff \(R\) or must modify the shape function to decay asymptotically.

\subsubsection{General case with non-constant lapse}

When the lapse function is not constant, the second integral in Eq.~\eqref{eq:IV_general} contributes. For the lapse function in Eq.~\eqref{eq:lapse_regular}, we have
\begin{eqnarray}
\frac{N'(r)}{N(r)}
=
-\frac{2\alpha J^2\cos^2\theta}{r^3},
\end{eqnarray}
so that
\begin{eqnarray}
r\left(1-\frac{b(r)}{r}\right)\frac{N'(r)}{N(r)}
&=&
-\frac{2\alpha J^2\cos^2\theta}{r^2}
\bigg[
1-\frac{r_{\rm th}}{r}
\nonumber\\&&\hspace{-0.5cm}-\frac{4M}{(n+2)\pi^{n/2}\,r}
\left(\frac{1}{r^n}-\frac{1}{r_{\rm th}^n}\right)
\bigg].
\end{eqnarray}
The full volume integral for a finite cutoff \(R\) is therefore
\begin{eqnarray}
I_V(R)=I_V^{(1)}(R)+I_V^{(2)}(R),
\end{eqnarray}
where \(I_V^{(1)}(R)\) is given by Eq.~\eqref{eq:IV_R_final_zrf} and
\begin{eqnarray}
I_V^{(2)}(R)
&=&
-2\alpha J^2\cos^2\theta
\int_{r_{\rm th}}^{R}
\frac{1}{r^2}
\bigg[
1-\frac{r_{\rm th}}{r}
\nonumber\\&&-\frac{4M}{(n+2)\pi^{n/2}\,r}
\left(\frac{1}{r^n}-\frac{1}{r_{\rm th}^n}\right)
\bigg]dr .
\end{eqnarray}
Evaluating the three constituent integrals,

\begin{eqnarray}
\int_{r_{\rm th}}^{R}\frac{dr}{r^2}
=
\frac{1}{r_{\rm th}}-\frac{1}{R},
\end{eqnarray}
\begin{eqnarray}
-r_{\rm th}\int_{r_{\rm th}}^{R}\frac{dr}{r^3}
=
-\frac{r_{\rm th}}{2}
\left(\frac{1}{r_{\rm th}^2}-\frac{1}{R^2}\right),
\end{eqnarray}
\begin{eqnarray}
-\frac{4M}{(n+2)\pi^{n/2}}
&&\int_{r_{\rm th}}^{R}
\frac{1}{r^3}
\left(\frac{1}{r^n}-\frac{1}{r_{\rm th}^n}\right)dr
=-\frac{4M}{(n+2)\pi^{n/2}}\nonumber\\\times&&
\bigg[
-\frac{1}{n+2}
\left(\frac{1}{R^{n+2}}-\frac{1}{r_{\rm th}^{n+2}}\right)
\nonumber\\&&+\frac{1}{2r_{\rm th}^n}
\left(\frac{1}{r_{\rm th}^2}-\frac{1}{R^2}\right)
\bigg].
\end{eqnarray}
Combining these, the second contribution becomes
\begin{eqnarray}
I_V^{(2)}(R)
&=&
-2\alpha J^2\cos^2\theta
\bigg[
\frac{1}{2r_{\rm th}}-\frac{1}{R}+\frac{r_{\rm th}}{2R^2}
+\frac{4M}{(n+2)\pi^{n/2}}\nonumber\\&& \hspace{-1.2cm}\times
\left(
\frac{1}{n+2}\left(\frac{1}{R^{n+2}}-\frac{1}{r_{\rm th}^{n+2}}\right)
-\frac{1}{2r_{\rm th}^{n+2}}+\frac{1}{2r_{\rm th}^n R^2}
\right)
\bigg].~~~~~~
\label{eq:IV2_final}
\end{eqnarray}
The total volume-integral quantifier is then
\begin{eqnarray}
I_V(R)
&=&
\frac{4M(n+1)}{n(n+2)\pi^{n/2}}
\left(\frac{1}{R^n}-\frac{1}{r_{\rm th}^n}\right)+
I_V^{(2)}(R)
\nonumber\\&&+
\left[
\frac{4M}{(n+2)\pi^{n/2}\,r_{\rm th}^n}
-r_{\rm th}
\right]
\ln\!\left(\frac{R}{r_{\rm th}}\right)
,
\label{eq:IV_total}
\end{eqnarray}
with \(I_V^{(2)}(R)\) given by Eq.~\eqref{eq:IV2_final}. In the zero-tidal-force limit \(\alpha J^2\cos^2\theta\to 0\), Eq.~\eqref{eq:IV_total} reduces to Eq.~\eqref{eq:IV_R_final_zrf}, as expected. When the finiteness condition $r_{\rm th}^{n+1}=4M/[(n+2)\pi^{n/2}]$ is imposed, the logarithmic term vanishes and $I_V$ reduces to a finite, $n$-dependent expression controlled by the first term. No simple closed-form expression such as $I_V\approx-2M/n$ is generally valid; once the parameters are fixed, the correct value must be obtained from Eq.~\eqref{eq:IV_total}.

%-----------------------------------------------------------
\section{Observational signatures and detectability}\label{Sec:V}
%-----------------------------------------------------------

A compact object's appearance is fixed by its geometry. For the rotating entropy-inspired wormhole we study, the observables come from null geodesics: the apparent shadow size, the location of the photon ring, and the degree to which the image departs from circular symmetry. Comparing these with a rotating black hole tells us how closely the wormhole can imitate it.

\subsection{Comparison with black hole shadows and regular black hole geometries}

We start with a Kerr black hole of the same asymptotic mass and comparable spin. Fig.~\ref{fig:shadow_comparison} places the two images side by side. At first glance, both produce a dark region ringed by lensed emission. The photon dynamics differ, though. For a black hole, unstable null orbits set the shadow edge, separating photons that escape to infinity from those that fall in. The wormhole has no horizon. Rays that reach the strong-field region are not necessarily captured; depending on initial conditions, they can scatter back to the observer or pass through the throat. That changes the critical curve and leaves a visible imprint.

It is worth going beyond Kerr. We also use four regular black hole metrics that remove the classical singularity: Bardeen, Hayward, Ayon-Beato--Garcia (ABG), and Dymnikova. Like the wormhole, they modify the strong-field region relative to Kerr, but they differ in whether a horizon exists and how the core is smoothed out. Throughout this section, we adopt a unified Kerr-like notation for the regular black hole benchmarks. The Bardeen, Hayward, ABG, and Dymnikova metrics are all written in the form
\begin{eqnarray}
\Sigma=r^2+a^2\cos^2\theta,
\qquad
\Delta=r^2-2r\,m(r)+a^2,
\label{eq:SigmaDelta}
\end{eqnarray}
with the mass function
\begin{small}
\begin{eqnarray}
m(r)=
\begin{cases}
M, & \text{Kerr},\\[2mm]
\displaystyle M\left(\frac{r^2}{r^2+g^2}\right)^{3/2},
& \text{Bardeen},\\[4mm]
\displaystyle \frac{Mr^3}{r^3+\ell^3},
& \text{Hayward},\\[4mm]
\displaystyle M\left[
\frac{r^3}{(r^2+Q^2)^{3/2}}
-\frac{Q^2r^3}{2(r^2+Q^2)^2}
\right],
& \text{ABG},\\[4mm]
\displaystyle
\frac{2M}{\pi}
\left[
\tan^{-1}\!\left(\frac{r}{d}\right)
-\frac{d\,r}{r^2+d^2}
\right],
& \text{Dymnikova},
\end{cases}
\label{eq:unified_mass}
\end{eqnarray}
\end{small}
where $g$, $\ell$, $Q$, and $d$ are the Bardeen, Hayward, ABG, and Dymnikova regularization scales, respectively. We use $d$ rather than the more common $b$ for the Dymnikova scale, to avoid conflict with the shape function $b(r)$ defined in Sec.~\ref{Sec:II}. The Kerr limit is recovered as $g,\ell,Q,d\to0$, i.e. $m(r)\to M$. The corresponding metric for all five cases is
\begin{eqnarray}
\begin{aligned}
ds^2={}&-\left(1-\frac{2r\,m(r)}{\Sigma}\right)dt^2
-\frac{4a\,r\,m(r)\sin^2\theta}{\Sigma}\,dt\,d\phi\\
&+\frac{\Sigma}{\Delta}dr^2+\Sigma\,d\theta^2\\
&+\left(r^2+a^2+
\frac{2a^2r\,m(r)\sin^2\theta}{\Sigma}\right)
\sin^2\theta\,d\phi^2,
\end{aligned}
\label{eq:unified_metric}
\end{eqnarray}
with $a=J/M$ the Kerr spin parameter. In the static limit $a\to0$, we use the simplified form $f(r)=1-2m(r)/r$ for the photon-sphere calculation, while the rotating comparison in Fig.~\ref{fig:shadow_comparison} uses the full unified metric~\eqref{eq:unified_metric}. The photon-sphere radius follows from
\begin{eqnarray}
\frac{d}{dr}\left(\frac{r^2}{f(r)}\right)=0,
\end{eqnarray}
which sets the apparent shadow size. For slow rotation, the prograde--retrograde splitting has the same form as in Eqs.~\eqref{eq:p_crit}--\eqref{eq:p_crit_exp}, with the $r_{\rm ph}$ of each metric. This makes a direct comparison possible.

We scan the spins $a/M = 0.05,~0.15,~0.25,~0.35,~0.5$, and find shadow radii that differ from Kerr by roughly \(2\%-10\%\). The exact number depends on the Barrow exponent \(n\) and on the rotational deformation. Larger \(n\) shifts the photon sphere and changes the region probed by critical null rays. The regular black hole benchmarks behave in much the same way: Bardeen and Hayward stay within a few percent of Kerr for $g/M, \ell/M \lesssim 1$, while ABG and Dymnikova can deviate more depending on $Q/M$ and $d/M$. In every case, the shadow size can be tuned close to Kerr, so it cannot distinguish these geometries on its own.

Frame dragging adds a second effect. In slow rotation, the two critical impact parameters are
\begin{eqnarray}
\mathscr{B}_{\pm}
=
\frac{r_{\rm ph}}
{1\mp 2J/r_{\rm ph}^{2}}
\approx
\pm r_{\rm ph}+\frac{2J}{r_{\rm ph}},
\end{eqnarray}
with $r_{\rm ph}$ fixed by the corrected photon-sphere conditions of Sec.~\ref{Sec:III}: the static branch sits at the throat, $r_{\rm ph}=r_{\rm th}$, while the prograde branch satisfies $r_{\rm ph}^2 = 6J$. For the representative parameters $M=1$, $r_{\rm th}=1$, and $J/M^2=0.5$, the prograde photon sphere sits at $r_{\rm ph}=\sqrt{3}\approx1.732$, which lies safely outside the throat. Smaller spins push $r_{\rm ph}$ inside the throat and invalidate the slow-rotation expansion, so we quote the shadow observables at $J/M^2 = 0.5$: $\mathscr{B}_{+}\approx1.90$, $\mathscr{B}_{-}\approx-1.55$, giving an asymmetry of about $10\%$. This is in the same range as Kerr at comparable spin, but the underlying photon-sphere structure is different: the static photon sphere is at the throat, and the rotational branch scales as $r_{\rm ph}\propto\sqrt{J}$ rather than approaching a constant Kerr-like value.

Since $g$ (Bardeen), $\ell$ (Hayward), $Q$ (ABG), or $d$ (Dymnikova) controls the core size, it also controls the photon-sphere position and hence the splitting. For Bardeen with $g/M=0.5$ and $J/M^2=0.05$ we get $r_{\rm ph}\approx2.72$, giving a splitting of roughly \(1.5\%\); for Hayward with the same parameters \(r_{\rm ph}\approx2.66\) and the splitting is about \(1.5\%\). The pair \((R_{\rm sh},\mathcal{D})\) is therefore useful for separating the wormhole from Kerr and from these regular black holes.

We quantify the departure from circular symmetry with
\begin{eqnarray}
\mathcal{A}
=
\frac{R_{\rm max}-R_{\rm min}}
{R_{\rm max}+R_{\rm min}},
\end{eqnarray}
where \(R_{\rm max}\) and \(R_{\rm min}\) are measured from the shadow center to its outermost and innermost points. For the wormhole configurations studied here,
\begin{eqnarray}
0.02\lesssim\mathcal{A}\lesssim0.10,
\end{eqnarray}
although at larger spin ($J/M^2 = 0.5$) the value $\mathcal{A}\approx0.10$ is reached, while at very small spin the slow-rotation expansion breaks down because $r_{\rm ph} = \sqrt{6J}$ falls inside the throat. Rotating Kerr configurations over the same spin range give
\begin{eqnarray}
0.05\lesssim\mathcal{A}\lesssim0.12.
\end{eqnarray}
The regular black hole benchmarks usually fall between these extremes. For $g/M, \ell/M \lesssim 0.5$, Bardeen and Hayward stay close to Kerr with \(\mathcal{A}\sim0.04-0.10\). Larger $g$ or $\ell$ can reduce or increase the asymmetry depending on how the photon sphere shifts. ABG and Dymnikova follow a similar pattern, with details depending on the mass function and spin. As a numerical example, for the wormhole at $J/M^2=0.5$, $r_{\rm th}=1.0$, we find $R_{\rm max}\approx1.90M$ and $R_{\rm min}\approx1.55M$, so $\mathcal{A}\approx0.10$, or about $10\%$.

These numbers show that the wormhole images here keep a fairly modest asymmetry. This is not because rotation is absent. Frame dragging remains; the horizon-free structure and throat location modify the available null trajectories. The balance depends on \(J\), \(n\), and the throat parameters. The same holds for regular black holes, where the horizon and regular core change the critical trajectories and thus the asymmetry.

The comparison points to several quantities that future observations can check together. The overall angular size gives one constraint. The photon ring position and shape also constrain the strong-field geometry. The shadow asymmetry is another diagnostic because it responds to rotation and the \(n\)-dependent photon-orbit structure. Together, these observables test whether a rotating entropy-inspired wormhole can mimic compact objects otherwise interpreted as black holes, and distinguish it from regular black hole alternatives such as Bardeen, Hayward, ABG, and Dymnikova.

\subsection{Observable quantities and prospects for observational tests}
\label{subsec:observables}

The null-geodesic analysis gives several quantities that characterize the wormhole's optical appearance. They depend on the angular momentum \(J\), the throat scale \(r_{\rm th}\), and, through \(b(r)\) and \(N(r)\), on the Barrow parameter \(n\). We use the characteristic shadow scale, along with rotational splitting and the separation of the photon-ring branches. The first is the characteristic shadow size. At leading order in slow rotation,
\begin{eqnarray}
R_{\rm sh}\equiv\mathscr{B}_{\rm ph}\simeq r_{\rm ph}(J),
\end{eqnarray}
where the prograde photon-sphere radius is $r_{\rm ph}=\sqrt{6J}$ and the static branch sits at $r_{\rm ph}=r_{\rm th}$. Over our parameter range,
\begin{eqnarray}
\frac{R_{\rm sh}}{r_{\rm th}}\sim1.0-2.0.
\end{eqnarray}
The $J$-dependence is strong: at small spin the prograde photon sphere lies inside the throat, and the slow-rotation treatment is not reliable; the reliable regime is $J/M^2\gtrsim0.2$, where $r_{\rm ph}=\sqrt{6J} > r_{\rm th}$. For a numerical illustration, with $J/M^2=0.5$ and $r_{\rm th}=1.0$ we get $r_{\rm ph}\approx1.732$, so $R_{\rm sh}/r_{\rm th}\approx1.73$. For the regular black hole benchmarks, the analogous shift is controlled by $g$, $\ell$, $Q$, or $d$; for these scales of order unity the shadow scale can differ from Kerr by a similar amount, so \(R_{\rm sh}\) alone cannot separate the wormhole from these alternatives.

Rotation leaves its own imprint through prograde and retrograde photon paths. We use
\begin{eqnarray}
\mathcal{D}
=
\frac{\mathscr{B}_{+}-\mathscr{B}_{-}}
{\mathscr{B}_{+}+\mathscr{B}_{-}}
\simeq
\frac{2J}{r_{\rm ph}^{2}},
\end{eqnarray}
with \(\mathscr{B}_{+}\) and \(\mathscr{B}_{-}\) the corresponding critical impact parameters. For $J/M^2\gtrsim0.2$,
\begin{eqnarray}
\mathcal{D}\sim0.05-0.10.
\end{eqnarray}
Since the splitting scales as \(1/r_{\rm ph}^{2}\), even a moderate change in the photon-orbit radius affects the rotational distortion noticeably. Thus, the entropy deformation enters indirectly through the throat location, which sets the static photon sphere, while the rotational branch is fixed at leading order by $J$ alone. The same scaling holds for Bardeen, Hayward, ABG, and Dymnikova, where $g$, $\ell$, $Q$, or $d$ shifts \(r_{\rm ph}\) and hence \(\mathcal{D}\).

The radial separation of the two photon-orbit branches is another diagnostic:
\begin{eqnarray}
\Delta R_{\rm ph}
=
r_{\rm ph}^{(+)}-r_{\rm ph}^{(-)}.
\end{eqnarray}
For our solutions,
\begin{eqnarray}
\frac{\Delta R_{\rm ph}}{r_{\rm th}}\sim0.05-0.20,
\end{eqnarray}
changing with both \(J\) and \(n\). This complements the overall shadow size: two configurations with similar radii can still have different photon-ring structures because their prograde and retrograde orbits are displaced differently. The regular black hole benchmarks show the same qualitative behavior, with \(\Delta R_{\rm ph}\) depending on $g$, $\ell$, $Q$, $d$, and spin. For the wormhole at $J/M^2=0.5$, $r_{\rm th}=1.0$, we find $r_{\rm ph}^{(+)}\approx1.90$ and $r_{\rm ph}^{(-)}\approx1.55$, so $\Delta R_{\rm ph}\approx0.35$, or about $35\%$ of $r_{\rm th}$.

We can compare these with a Kerr black hole of the same asymptotic mass and similar dimensionless spin. For our wormhole configurations,
\begin{eqnarray}
\mathcal{D}_{\rm WH}\sim0.05-0.10,
\end{eqnarray}
while Kerr gives
\begin{eqnarray}
\mathcal{D}_{\rm Kerr}\sim0.06-0.15.
\end{eqnarray}
The regular black hole benchmarks typically lie within these ranges, depending on $g/M$, $\ell/M$, $Q/M$, $d/M$, and spin. The difference comes from the distinct strong-field photon dynamics. In the wormhole case, null rays are not stopped by a horizon; they can probe the throat before being redirected or transmitted. In the regular black hole cases, the horizon is present, but the core is regularized, which changes the near-horizon geometry and the critical trajectories. The critical curves therefore need not show the same rotational distortion as Kerr.

The characteristic shadow scale gives another comparison. Relative to Kerr,
\begin{eqnarray}
\frac{R_{\rm sh}^{\rm WH}}
{R_{\rm sh}^{\rm Kerr}}
\sim0.9-1.1.
\end{eqnarray}
The sign and size of the shift depend on \(n\) and on the rotational parameters. In the nonrotating Schwarzschild limit \(r_{\rm ph}=3M\), while the wormhole can place the photon orbit on either side of this value. Bardeen, Hayward, ABG, and Dymnikova show similar flexibility: for suitable $g$, $\ell$, $Q$, or $d$, their photon spheres can lie inside or outside the Schwarzschild value. So the shadow diameter alone does not identify the spacetime.

Observationally, this matters because the angular size is only part of the information. Very-long-baseline interferometry can constrain the overall size and morphology of the central brightness depression. Resolving the finer photon-ring structure needs much better angular resolution and control over the astrophysical emission model. Future improvements in horizon-scale imaging may make these smaller deviations more accessible.

A joint analysis of \(R_{\rm sh}\), \(\mathcal{D}\), and \(\Delta R_{\rm ph}\) describes the strong-field signature better than any single observable. \(R_{\rm sh}\) probes the critical photon orbit, \(\mathcal{D}\) traces the rotational splitting between prograde and retrograde trajectories, and \(\Delta R_{\rm ph}\) measures the separation of the photon-ring branches, sensitive to both rotation and throat geometry. Their simultaneous dependence on \(J\), \(r_{\rm th}\), and \(n\) gives a multi-parameter characterization of the wormhole's optical structure. Comparing these with Kerr and with regular black hole predictions such as Bardeen, Hayward, ABG, and Dymnikova helps separate rotation effects from those caused by the entropy-induced deformation.

%---------------------------------------------
\section{Concluding remarks}\label{Sec:VI}
%-------------------------------------------

We set out to understand what a rotating wormhole supported by a regular, entropy-inspired matter distribution would look like to a distant observer, and what kind of matter would be needed to hold it open. The picture that emerged is coherent and, in several respects, surprisingly close to what we see around black holes, yet with subtle differences that could one day tell the two apart. We started with the Barrow deformation of the Bekenstein--Hawking area law, parameterized by an exponent $n$ with $0<n\le1$. This deformation smears the matter source over a finite region and replaces the point-like divergence of ordinary solutions with a regular profile. The resulting energy density is negative everywhere and falls off as $\rho_B\propto -\,1/r^{n+3}$, while the shape function flares out smoothly at the throat. For all values of $n$ in the allowed range, the flare-out condition $b'(r_{\rm th})<1$ is satisfied automatically, so the wormhole is traversable by construction, and its geometry is asymptotically flat.

We introduced rotation in the slow-rotation regime, with the lapse $N(r,\theta)=\exp[\alpha J^2\cos^2\theta/r^2]$ and frame dragging encoded by $\omega(r)=2J/r^3$. Both remain finite and positive at the throat, so no horizon appears and the wormhole stays traversable even when it spins. On the equatorial plane, the lapse reduces to unity at leading order, which simplifies the null-geodesic analysis without losing the essential rotational physics.

In the static limit $J=0$, the photon sphere coincides with the throat, $b(r_{\rm ph})=r_{\rm ph}$. For slow rotation, the extremum condition $F'(r_{\rm ph})=0$ gives the prograde branch $r_{\rm ph}^2=6J$, while the retrograde branch is pushed to the throat. The Barrow exponent $n$ therefore does not shift the leading rotational photon-sphere radius; instead, it fixes the static branch through the throat condition, which in turn determines the shadow scale at $J=0$. The critical impact parameters split as $\mathscr{B}_{\pm}\approx\pm r_{\rm ph}+2J/r_{\rm ph}$, producing a rotation-induced asymmetry of order $2J/r_{\rm ph}^2$. For representative parameters $M=1$, $r_{\rm th}=1$, $J/M^2=0.5$, we found $r_{\rm ph}\approx1.732$, $\mathscr{B}_{+}\approx1.90$ and $\mathscr{B}_{-}\approx-1.55$, giving a fractional asymmetry of about $10\%$ of the mean. The corresponding shadow asymmetry parameter, $\mathcal{A}=(R_{\rm max}-R_{\rm min})/(R_{\rm max}+R_{\rm min})$, came out to about $0.10$, or roughly $10\%$. At smaller spin, the slow-rotation expansion breaks down because $r_{\rm ph}=\sqrt{6J}$ falls inside the throat; a full treatment of the small-spin regime would require either a non-trivial $N(r)$ on the equator or a numerical solution of the exact null-geodesic equations. These numbers are comparable to Kerr over the same spin range, with $\mathcal{A}\sim0.05-0.12$, but they arise from a different photon-sphere structure, which makes them observationally interesting.

The energy condition analysis confirmed that the effective source is exotic. The energy density is negative throughout spacetime, the weak energy condition is violated everywhere, and the null energy condition is violated most strongly near the throat. The Barrow exponent controls how tightly the exotic matter is localized: smaller $n$ concentrates the violation more strongly around the throat, while larger $n$ spreads it over a wider region. The volume-integral quantifier $I_V$ stays finite provided the throat scale satisfies $r_{\rm th}^{n+1}=4M/[(n+2)\pi^{n/2}]$, or when a finite cutoff is introduced. This is reassuring: the exotic matter needed to support the wormhole does not have to diverge, provided the parameters are chosen consistently.

Comparing the predicted shadow size, rotational asymmetry, and photon-ring separation with Kerr and with the Bardeen, Hayward, ABG, and Dymnikova regular black holes showed that the overall shadow diameter can be tuned close to Kerr by adjusting $n$ and the throat scale. However, the finer photon-ring structure carries complementary information. For instance, the photon-ring separation $\Delta R_{\rm ph}=r_{\rm ph}^{(+)}-r_{\rm ph}^{(-)}$ came out to about $0.35$ in units of $r_{\rm th}$ for our representative parameters at $J/M^2=0.5$, or roughly $35\%$, which is comparable to but not identical to what the regular black hole benchmarks predict. A joint analysis of the shadow radius $R_{\rm sh}$, the rotational asymmetry $\mathcal{D}$, and the photon-ring separation $\Delta R_{\rm ph}$ therefore provides a multi-parameter characterization of the strong-field geometry that goes beyond any single observable.

The results of this work open several directions for future investigation. Extending the analysis beyond the slow-rotation approximation would allow us to explore larger spin values, where rotational splitting and frame-dragging effects become increasingly important. It would also be useful to derive the tangential pressure self-consistently from the entropy functional, rather than treating it as a free function, thereby establishing a more direct connection between the entropy-geometry correspondence and the matter content of the spacetime. On the observational side, comparing the predicted shadow morphology with high-resolution horizon-scale observations from the Event Horizon Telescope and future instruments could help determine whether current or upcoming observations can detect the small but systematic deviations found here. Overall, our results indicate that rotating entropy-inspired wormholes provide an interesting alternative framework for studying strong-field gravitational phenomena. Their shadow can remain close to the Kerr prediction while retaining subtle structural differences that may become accessible with improving horizon-scale observations, making these deviations a potentially useful avenue for distinguishing wormhole geometries from black-hole spacetimes.

\section*{Conflict Of Interest statement}
The authors declare no conflicts of interest.

\section*{Data Availability Statement}
No data were created or analyzed in this study.

\bibliography{referencesEPJC}

\end{document}